\documentclass[preprintnumbers,prd,onecolumn,floatfix,superscriptaddress,nofootinbib]{revtex4-2}

\usepackage{setspace}
\usepackage{graphicx}
\usepackage{subfig}
\usepackage{epsfig}
\usepackage{bm}
\usepackage{amssymb}
\usepackage{float}
\usepackage{amsmath}
\usepackage{dcolumn}
\usepackage[colorlinks]{hyperref}
\usepackage[usenames,dvipsnames]{color}
\usepackage{enumitem}

\hypersetup{ breaklinks=true, pdfstartview={FitH}, colorlinks=true, linkcolor=blue, citecolor=red, filecolor=magenta, urlcolor=blue, anchorcolor=green, linktocpage=true }

\def\doi{http://doi.org}

\newcommand{\be}{\begin{equation}}
\newcommand{\ee}{\end{equation}}
\newcommand{\beano}{\begin{eqnarray*}}
\newcommand{\eeano}{\end{eqnarray*}}
\newcommand{\ba}{\begin{eqnarray}}
\newcommand{\ea}{\end{eqnarray}}

\begin{document}

\title{ The consequence of higher-order curvature-based constraints on $ f(R, L_m) $ gravity}
\author{J. K. Singh}
\email{jksingh@nsut.ac.in}
\affiliation{Department of Mathematics, Netaji Subhas University of Technology, New Delhi-110078, India}
\author{Shaily}
\email{shailytyagi.iitkgp@gmail.com}
\affiliation{Department of Mathematics, Netaji Subhas University of Technology, New Delhi-110078, India}
\affiliation{School of Computer Science Engineering and Technology, Bennett University, Greater Noida, India}
\author{Akanksha Singh}
\email{akanksha.ma19@nsut.ac.in}  
\affiliation{Department of Mathematics, Netaji Subhas University of Technology, New Delhi-110078, India}
\author{Harshna Balhara}
\email{harshna.ma19@nsut.ac.in}  
\affiliation{Department of Mathematics, Netaji Subhas University of Technology, New Delhi-110078, India}

\author{Joao R. L. Santos}
\email{joaorafael@df.ufcg.edu.br}
\affiliation{UFCG-Universidade Federal de Campina Grande-Unidade Academica de Fisica, 58429-900 Campina Grande, PB, Brazil}

\begin{abstract}
\begin{singlespace}

In this investigation, we perform an observational statistical analysis in the theory of $ f(R, L_m) $ gravity. The proposed theoretical model is based on the Ricci scalar's non-linear contribution. We use a distinct parameterization for the deceleration parameter and constrain the model parameters by using various observational data. To determine the best-fit model for the cosmological parameters, we use different observational datasets such as the Hubble Space Telescope, the Pantheon Supernova Survey, the Gold dataset, the Gamma-Ray Burst (GRB), and the Baryon Acoustic Oscillations (BAO). Furthermore, we study the late-time cosmic evolution of the Universe in detail and examine the implications of the constraint values on cosmological parameters. Additionally, we conduct a thorough comparison with the standard cosmological model $ \Lambda $CDM and other standard models obtained by Odintsov et al. \cite{Odintsov:2023cli, Odintsov:2024lid} to examine the validity of our proposed model in the low-redshift regimes. Finally, we find that the proposed model encapsulates an intriguing transition from early deceleration at high redshift to acceleration at low redshift, a quintessence dark energy scenario, and convergence towards the well-established $ \Lambda $CDM model in late-time Universe's evolution.

\end{singlespace}
\end{abstract}
 
\maketitle
PACS numbers: {04.20.-q, 04.50.Kd, 98.80.Es}\\

Keywords: $ f(R, L_m) $ gravity, Observational analysis, FLRW metric, dark energy

\section{Introduction} \label{I}

\qquad In the pursuit of unraveling the cosmic mysteries that govern the fate of our Universe, the $ 2011 $ Nobel Prize in Physics acknowledged Saul Perlmutter, Brian Schmidt, and Adam Reiss for their relevant contributions in describing the dark energy era. Their revelation of the accelerating expansion of the Cosmos, in direct contrast to the anticipated deceleration under the gravitational influence of matter, is a milestone in cosmology. This discovery emerged from meticulous comparisons between the luminosity of distant supernovae and their nearby counterparts. The observation that distant supernovae appeared approximately $ 25\% $ fainter, indicative of their greater distance, provided strong evidence for the Universe's acceleration. This picture becomes clear as high-quality data is obtained from several observational surveys, collectively contributing to the framework of precision cosmology. Current investigations, including the scrutiny of type Ia supernovae \cite{SupernovaSearchTeam:1998fmf, SupernovaCosmologyProject:1998vns}, explorations of cosmic microwave background radiation anisotropies \cite{WMAP:2003xez, Caldwell:2003hz}, measurements of the matter power-spectrum \cite{2dFGRS:2005yhx, Koivisto:2005mm, Daniel:2008et}, and noteworthy contributions from endeavors such as the Sloan Digital Sky Survey \cite{SupernovaSearchTeam:2004lze}, Wilkinson Microwave Anisotropy Probe \cite{WMAP:2003elm}, and Baryonic Acoustic Oscillations \cite{SDSS:2005xqv, SDSS:2009ocz} collectively underscore and support the observed acceleration dynamics \cite{Jassal:2005qc}. Besides, we can highlight several theoretical proposals searching for the nature of dark energy \cite{Padmanabhan:2006ag, Copeland:2006wr, Padmanabhan:2007xy, Durrer:2007re, Capozziello:2011et, Clifton:2011jh, Bamba:2012cp, Will:2014kxa, Joyce:2014kja, Cai:2015emx, Bahamonde:2021gfp, Langlois:2018dxi, Frusciante:2019xia, Arai:2022ilw, Odintsov:2023weg}.

In cosmology, the accelerated expansion of the Universe has sparked inquiries into the fundamental nature of cosmic evolution. General relativity (GR) contends with this cosmic problem by introducing the concept of dark energy (DE), an energy component characterized by a large negative pressure. At the forefront of DE representations lies the cosmological constant $\Lambda$, a parameter with a measured value that defies conventional expectations by spanning $ 55 $ orders of magnitude \cite{Carroll:2000fy, Peebles:2002gy, Carroll:2003wy}. While $ \Lambda $ stands as the most accepted description of dark energy, several other possible candidates emerge when considering new theories of gravity, indicating a preference for dynamical dark energy linked with a scalar field \cite{Wetterich:1987fm, Ratra:1987rm, Caldwell:1997ii, Armendariz-Picon:1999hyi, Armendariz-Picon:2000nqq, Armendariz-Picon:2000ulo, Mersini-Houghton:2001cwp, Caldwell:1999ew, Carroll:2003st, Sahni:1999gb, Parker:1999td}. Although most of these works have some challenges and drawbacks, such as the absence of an explicit fluid or substance driving the Universe's expansion, these approaches fundamentally attribute this phenomenon to the principles of gravitational physics \cite{Deffayet:2001pu, Freese:2002sq, Ahmed:2002mj, Arkani-Hamed:2002ukf, Dvali:2003rk}. One of the approaches to dealing with this challenge is the compatibility of extended modified gravity with observational outcomes, as we can see, for instance, in the following references: \cite{deHaro:2023lbq, Singh:2018xjv, Singh:2022jue, Aviles:2014rma, delaCruz-Dombriz:2016bqh, Shaily:2024rjq, Capozziello:2019cav, Shaily:2024nmy, Balhara:2023mgj, Singh:2024ckh, Shaily:2024xho}.

At the solar system's scale, general relativity is a keystone in theoretical physics. Within this framework, the Einstein field equations unfold, serving as a powerful tool for unraveling the intricacies of the Cosmos. These equations, derived from the Einstein-Hilbert action $ S=\frac{1}{16\pi G} \int (R+L_m)\sqrt{-g}d^4x $, capture the essence of gravitational interactions by incorporating the Ricci scalar $ R $ and the matter Lagrangian $ L_m $. The Einstein field equations are encapsulated in the form $ R_{ij}-\frac{1}{2}R g_{ij}=\frac{8 \pi G}{c^4} T_{ij} $, where $ T_{ij} $ stands for the energy-momentum tensor. A crucial mathematical prerequisite is $ \nabla^i T_{ij}=0 $, enforcing energy conservation. For large-scale Cosmos, numerical simulations are performed through GR, verifying its validity with the help of observational data techniques. In $ f(R) $ gravity, several models permit the local tests and link the inflation with dark energy \cite{Nojiri:2007as, Nojiri:2007cq, Cognola:2007zu}. Moreover, in some models, the galactic nature of massive test particles is discussed without any use of dark matter \cite{Capozziello:2006ph, Borowiec:2006qr, Martins:2007uf, Boehmer:2007kx}. Among the new proposals to describe gravity, we can highlight those that feature an explicit coupling between geometry and matter, as we can see in the so-called $ f(R, L_m) $ theory of gravity \cite{Sotiriou:2008rp, DeFelice:2010aj, Bertolami:2007gv, Nojiri:2010wj, Nojiri:2017ncd, Jana:2023djt, Maurya:2023lxt, Pradhan:2022msm}. Due to this coupling, we observe a non-geodesic motion for massive particles, accompanied by a force perpendicular to the four velocities. In both geometry and matter, such non-minimal couplings have different astrophysical and cosmological implications, which have been investigated in several studies \cite{Allemandi:2005qs, Nojiri:2004bi, Bertolami:2008zh, Harko:2008qz, Bertolami:2009cd, Solanki:2023onp, Jaybhaye:2023lgr, Nojiri:2006ri}. Among such approaches, Singh et al. discussed a cosmological model with higher-order curvature terms of the Ricci scalar and the Gauss-Bonnet invariant and performed the perturbation analysis on the Hubble parameter to check the stability of the model \cite{Singh:2024aml}.

Besides the previous works, we also highlight the studies carried out by Harko on a generalized gravity model in which an arbitrary function $ f(R, L_m) $ is taken as $ \frac{1}{2}f_1(R)+G(L_m)f_2(R) $, where $ f_1(R) $ and $ f_2(R) $ are functions of Ricci scalar and $ G(L_m) $ is any random function representing the density of the matter Lagrangian \cite{Harko:2008qz}. Furthermore, Kavya et al. considered a non-linear $ f(R, L_m) $ gravity within locally rotational symmetric homogeneous Bianchi-I space-time and analyzed the behavior of some cosmological parameters \cite{Kavya:2022dam}. Jaybhaye et al. considered a non-linear $ f(R, L_m) $ model, focusing on assessing model stability through observational constraints and linear perturbation analysis \cite{Jaybhaye:2022gxq}. Lobato et al. investigated neutron stars in $ f(R, L_m) $ gravity, explicitly considering the coupling between the gravitational and particle fields \cite{Lobato:2021ehf}. Gon\c{c}alves et al. also considered a geometry-matter coupling, the Big Bang singularity, and cosmic acceleration \cite{Goncalves:2021vim}. Harko et al. proposed a compelling model featuring exponential dependence of the gravitational field action on the Hilbert-Einstein Lagrange density, demonstrating its viability across various matter types, including vector fields, scalar fields, and Yang-Mills fields \cite{Harko:2010mv}. Wang et al. \cite{Wang:2012rw} contributed to the theoretical landscape by deriving energy conditions in $ f(R, L_m) $ gravity, exploring arbitrary coupling scenarios, non-minimal coupling, and the absence of coupling between geometry and matter. Singh et al. \cite{Singh:2022ptu} discussed the expanding nature of the Universe in the $ f(R, L_m) $ gravity theory, using the non-minimal coupling between matter and the linear curvature.

So, motivated by these works, our present manuscript approaches the $ f(R, L_m) $ using a different path based on a distinctive parametrization of the deceleration parameter $ q $. The observational part of the methodology of this paper uses $ 77 $ data points of the observational Hubble dataset, $ 1048 $ data points of the Pantheon compilation, $ 162 $ points of the Gamma-Ray Burst dataset, $ 185 $ points of the Gold dataset, and $ 6 $ points of the BAO dataset to constrain a non-minimal coupling between a non-linear Ricci scalar and the non-linear Lagrangian density for the matter. The outcomes of the model derived for the best-fitted values of the parameters yield an Equation of State (EoS) parameter $ \omega $ safe from the phantom zone and compatible with the present and late time eras of the Universe, which is favorable to standard cosmology. 

The key aspects of our research are organized in the following nutshell: In Sec. \ref{sec:2}, we present the Einstein field equations (EFEs) for $ f(R, L_m) $. We introduce a specific parametrization of the deceleration parameter to solve these field equations, providing a robust foundation for our subsequent analyses and discussions. Sec. \ref{sec:3} serves as a focal point where we constrain the model parameters, a critical step toward obtaining the best-fit values using the recent Observational Hubble dataset (OHD) ($77$ points), the Pantheon dataset ($ 1048 $ points), the Gold dataset ($ 185 $ points), the Gamma-Ray Burst ($ 162 $ points), and the BAO ($ 6 $ points). In Sec. \ref{sec:4}, we discuss the model's physical features and present the statefinder diagnostic technique in detail. Finally, we approach the cosmic consequences of our model in Sec. \ref{sec:5}.

\section{FLRW model in $ f(R, L_m) $ gravity} \label{sec:2}
For the modified theory of gravity $ f(R, L_m) $, the matter Lagrangian $ L_m $ is directly incorporated into the gravitational action of the theory. The Einstein field equations corresponding to this alternative gravitational theory are derived from the action stated as \cite{Harko:2010mv}
\begin{equation}\label{1}
S=\int f(R,L_m)\sqrt{-g}d^4x,
\end{equation}
where $ f(R, L_m) $ emerges as an arbitrary function intertwining the Ricci scalar $ R $ and the Lagrangian density to matter $ L_m $, which is defined as $ f(R, L_m)=F_1(R)+2 F_2(L_m) $. From the action mentioned in Eq. (\ref{1}), the Einstein field equations are written as
\begin{equation}\label{2}
R_{ij}-\frac{1}{2}R g_{ij}=\frac{8\pi G}{c^4} T_{ij},
\end{equation}
where the energy-momentum tensor incorporating contributions from the cosmic matter component is given as
\begin{equation}\label{3}
T_{ij}=-\frac{2}{\sqrt{-g}} \frac{\delta(\sqrt{-g} L_m)}{\delta g^{ij}}
\end{equation}
and the gravitational coupling constant, $ \kappa = \frac{8\pi G}{c^4} $, is normalized by setting it equal to $ 1 $.

As the dependency of the matter Lagrangian density $ L_m $ is solely on the metric tensor components $ g_{ij} $, excluding their derivatives, the following simplified expression is obtained
\begin{equation}\label{4}
T_{ij}=g_{ij}L_m - 2\frac{\partial L_m}{\partial g^{ij}}.
\end{equation}

Then, by taking the variation of Eq. (\ref{1}) with respect to the metric tensor $ g_{ij} $, we find the following equation
\begin{equation}\label{5}
\delta S= \int \Bigg[F_1^R(R)\delta R+ 2 F_2^{L_m}(L_m) \frac{\delta L_m}{\delta g^{ij}} \delta g^{ij}-\frac{1}{2} g_{ij} f(R,L_m) \delta g^{ij} \Bigg] \sqrt{-g} d^4x,
\end{equation}
where $ F_1^R=\frac{\partial F_1}{\partial R} $ and $ F_2^{L_m}=\frac{\partial F_2}{\partial L_m} $. Moreover, the variation of the Ricci scalar, $ \delta R= \delta(g^{ij}R_{ij}) $, results in
\begin{equation}\label{6}
\delta R=R_{ij}\delta g^{ij}+g^{ij}(\nabla_{k} \delta \Gamma^k_{ij}-\nabla_j \delta \Gamma^k_{ik} ),
\end{equation}
where $ \nabla_k $ is the covariant derivative and $ \delta \Gamma^k_{ij}=\frac{1}{2}g^{k\zeta}(\nabla_i \delta g_{j\zeta}+\nabla_j \delta g_{\zeta i}-\nabla_\zeta \delta g_{ij}) $ stands for the variation of the Christoffel symbol. Consequently, we can rewrite Eq. (\ref{6}) as
\begin{equation}\label{7}
\delta R= R_{ij}\delta g^{ij}+ g_{ij}\nabla_i \nabla^i \delta g^{ij}-\nabla_i \nabla_j \delta g^{ij}.
\end{equation}

Therefore, the Einstein field equations for $ f(R, L_m) $ gravity are calculated from Eq. (\ref{5}) as
\begin{equation}\label{8}
F_1^R R_{ij}+(g_{ij}\nabla_i \nabla^i - \nabla_i \nabla_j) F_1^R -\frac{1}{2}[f(R,L_m)- 2 F_2^{L_m} L_m]g_{ij}= F_2^{L_m} T_{ij}.
\end{equation}

In exploring the cosmic dynamics for the presented model, we turn our attention to a perfect fluid Universe characterized by $ T_{ij}=(\rho+p)u_i u_j+pg_{ij} $, where $ T_{ij} $, $ p $, and $ \rho $ denote energy-momentum tensor, isotropic pressure, and matter-energy density, respectively. Here, the four-velocity vector $ u_i $ takes the form $ u_i=(0,0,0,1) $ such that $ u^iu_i=-1 $. Here we also consider $ f(R, L_m) $ gravity in FLRW space-time, a choice that embraces the spatial homogeneity and isotropy inherent in our Universe. For our study, we adopted the following distinctive forms for the functions $ F_1(R) $ and $ F_2(L_m) $
\begin{equation}\label{9}
F_1(R)=R+\lambda R^2, ~~~~~F_2(L_m)= L_m^n.
\end{equation}

Let us now take a FLRW metric to define space-time evolution
\begin{equation}\label{10}
    ds^2 = -dt^2 + a^2(t)\sum_{i=1}^3 dx_i^2,
\end{equation}
where $ a(t) $ stands for the scale factor. For the present metric, the trace $ T $ of the energy-momentum tensor is given as $ T = -\rho+3 p $, and the Ricci scalar curvature is defined as $ R=6(2H^2+\dot{H}) $. Using Eq. (\ref{8}), the Einstein field equations within the framework of $f(R, L_m)$ gravity result into
\begin{equation}\label{11} 
3 H^2+108 \lambda \dot{H} H^2 - 60 \lambda H^4 - 18 \lambda \dot{H}^2 + 36 \lambda H \ddot{H}= \rho^n
\end{equation}
and
\begin{equation}\label{12}
-2 \dot{H} - 3 H^2 -54 \lambda \dot{H}^2 -156 \lambda \dot{H} H^2 - 12 \lambda \dddot{H} - 84 \lambda H \ddot{H}=n \rho^{n-1} p + (1-n) \rho^n,
\end{equation}
where $L_m=\rho$.

The Einstein field equations contain three variables $ H $, $ \rho $, and $ p $, which can be determined using some parametrization. Several methodologies in the literature yield solutions to field equations. These methods are specified as independent approaches, each exploring different DE models. Here, we are going to take a specific parametrization pattern for the deceleration parameter $ q $. Its direct relationship with the Hubble parameter is the primary motivation behind this assumption. A negative value for parameter $ q $ means the Universe is undergoing acceleration, while a positive value indicates the deceleration phase. When $ q=0 $, this means that we have a consistent and unchanging expansion rate. Inspired by \cite{Banerjee:2005ef}, we introduce the following form for the deceleration parameter
\begin{equation}\label{13}
q=-1+\frac{\alpha a^{\alpha-2}}{(1+a)^\alpha},
\end{equation}
where $a(t)$ is the scale factor and $\alpha$ is a real positive constant. To explore the cosmological solutions that allow for a direct correspondence of the model projections with the recent astronomical observations, we change the variable, rewriting our parameters in terms of the redshift $ z $, instead of time $ t $. The change of variable is based on the constraint $ a=\frac{a_0}{1+z} $, where the present value for the scale factor is $a_0=1$. Therefore, we can compute the following relation:
\begin{equation}\label{14}
\frac{d}{dt}=\frac{dz}{dt}\frac{d}{dz}=-(1+z)H(z)\frac{d}{dz}.
\end{equation}

Let us also impose that $ q=-1-\frac{\dot{H}}{H^2} $, which implies in
\begin{equation}\label{15}
q=-1+\frac{(1+z)}{H(z)}\frac{dH(z)}{dz}.
\end{equation}

Then, from Eqs. (\ref{13}) and (\ref{15}), the Hubble parameter is determined as
\begin{equation}\label{16}
H(z)= \frac{H_0}{3} e^{\frac{\alpha ^2}{(z+2)^{\alpha -1}}+\frac{\alpha -\alpha ^2}{(z+2)^{\alpha -2}}},
\end{equation}
where $\alpha$ and $H_0$ both are free model parameters and in the next section, their best-fit values are obtained. 

\section{Observational constraints} \label{sec:3}
\qquad This section involves an exploration of the cosmic fabric from a statistical perspective. Our discussions are based on five significant datasets representing recent statistical observations: the Observational Hubble dataset, the Pantheon dataset, the Gamma-Ray Burst dataset, the Gold dataset, and the Baryon Acoustic Oscillation dataset. In pursuit of a refined understanding of the model, we obtain the most probable values for $ \alpha $ and $ H_0 $. For this purpose, we minimize the value of the chi-square ($ \chi^2 $) and draw likelihoods using the Matplotlib library in Python with the Markov Chain Monte Carlo (MCMC) method.
    
\subsection{Observational Hubble dataset (OHD)}
\qquad We want to use observational datasets to constrain the best-fit values of model parameters $ \alpha $ and $ H_0 $. In this section, we will use the Hubble dataset with $ 77 $ data points acquired from measurements of the Hubble parameter in the redshift range $ 0.07\leq z \leq 2.42 $, including $ 31 $ points from the differential age method (DA method), and $ 46 $ points evaluated by using BAO data \cite{Shaily:2022enj, Singh:2023ryd, Singh:2023gxd, Singh:2024kez, Sharov:2018yvz}. We start our approach by minimizing the value of $ \chi^2 $, which is given as
\begin{equation}\label{17}
\chi_{OHD}^{2}(\alpha,H_0)=\sum\limits_{i=1}^{77} \frac{[H(\alpha,H_0,z_{i})-H_{obs}(z_{i})]^2}{\sigma_{z_i}^2},
\end{equation}
where $ H_{obs}(z_{i})$ and $ H(\alpha, H_0,z_{i}) $ indicate the observed and theoretical values of the Hubble parameter at redshift $ z_{i} $, respectively. The term $ \sigma_{z_{i}} $ shows the standard deviation associated with each $ H(z_i) $.

\begin{figure}\centering
	\subfloat[]{\label{fig:1a}\includegraphics[scale=0.8]{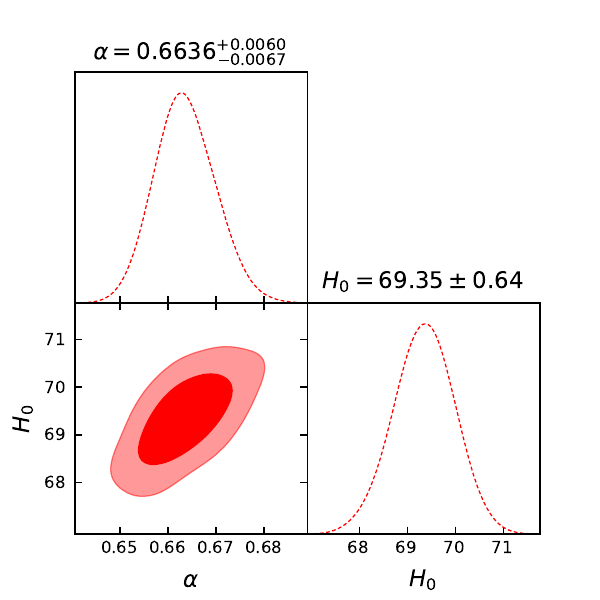}}\hfill
	\subfloat[]{\label{fig:1b}\includegraphics[scale=0.8]{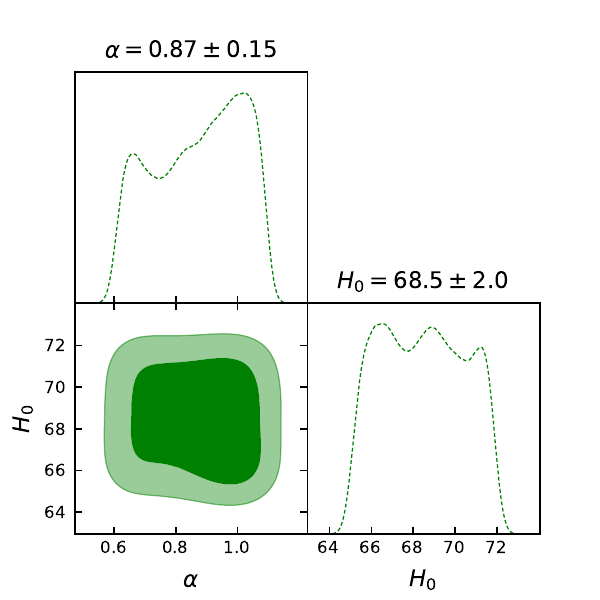}}\par
	\subfloat[]{\label{fig:1c}\includegraphics[scale=0.8]{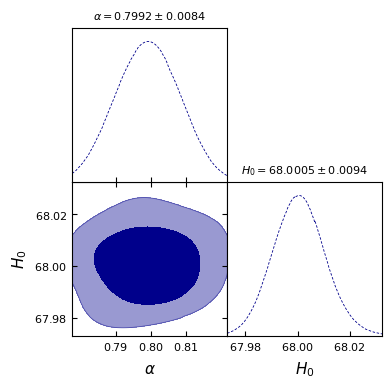}}
 \caption{ The likelihood contours for OHD, PGG (Pantheon+Gold+Gamma-Ray Burst), and HPGGB ($ H(z) $+Pantheon+Gold+Gamma-Ray Burst+BAO)} datasets with 1 $\sigma$ and 2 $\sigma$ confidence levels.
 \label{fig:1}
\end{figure}

\subsection{Pantheon, Gamma-Ray Burst and Gold datasets } 
\qquad The expansive Pantheon, Gamma-Ray Burst, and Gold datasets are an assembly of $ 1048, 162 $ and $ 185 $ points, which are obtained from various Supernovae type Ia (SNIa) surveys \cite{Pan-STARRS1:2017jku, Riess:1998dv, Jha:2005jg, Hicken:2009df, Contreras:2009nt, SDSS:2014irn, Izzo:2015vya, Demianski:2016dsa, Amati:2018tso} and cover the redshift ranges $ 0.01<z<2.26 $, $ 0.03351<z<9.3 $, and $ 0.0104<z<1.755 $, respectively. The SNIa data is used to explore the Universe's expansion rate. Therefore, the formula for the apparent magnitude is \cite{Alcaniz:2002fy, bogna/2024}
\begin{equation}\label{18}
    m(z)=M+5\log_{10}[d_L(z)]+5\log_{10}\Bigg[\frac{c/H_0}{Mpc}\Bigg]+25,
\end{equation}
where $ M $ signifies the absolute magnitude, $ d_L $ denotes the luminosity distance, $ c $ is the speed of light, and the Hubble constant is $ H_0=67.4 ~Km ~Sec^{-1}Mpc^{-1} $. In a flat Universe, the luminosity distance takes the form
\begin{equation}\label{19}
    d_L(z)=c (1+z) \int_0^z \frac{dz^*}{H(z^*,\zeta)}.
\end{equation}

Using the distance modulus definition, $ \mu = m-M $, we write the statistical $ \chi^2 $ for these datasets as 
\begin{equation}\label{20}
    \mathbf{ \chi_{PGG}^{2}(\alpha,H_0) }=\sum\limits_{i=1}^{1395}\left[ \frac{\mu_{the}(\alpha,H_0,z_{i})-\mu_{obs}(z_{i})}{\sigma _{\mu(z_{i})}}\right] ^2,
\end{equation}
where $ \mu_{the} $ is the theoretical distance modulus and $ \mu_{obs} $ represents the observed distance modulus. By minimizing $ \chi^2 $ and using the likelihood contour, we obtain the best-fit values for $ \alpha $ and $ H_0 $.
\begin{figure}\centering 
	\subfloat[]{\label{fig:2a}\includegraphics[scale=0.48]{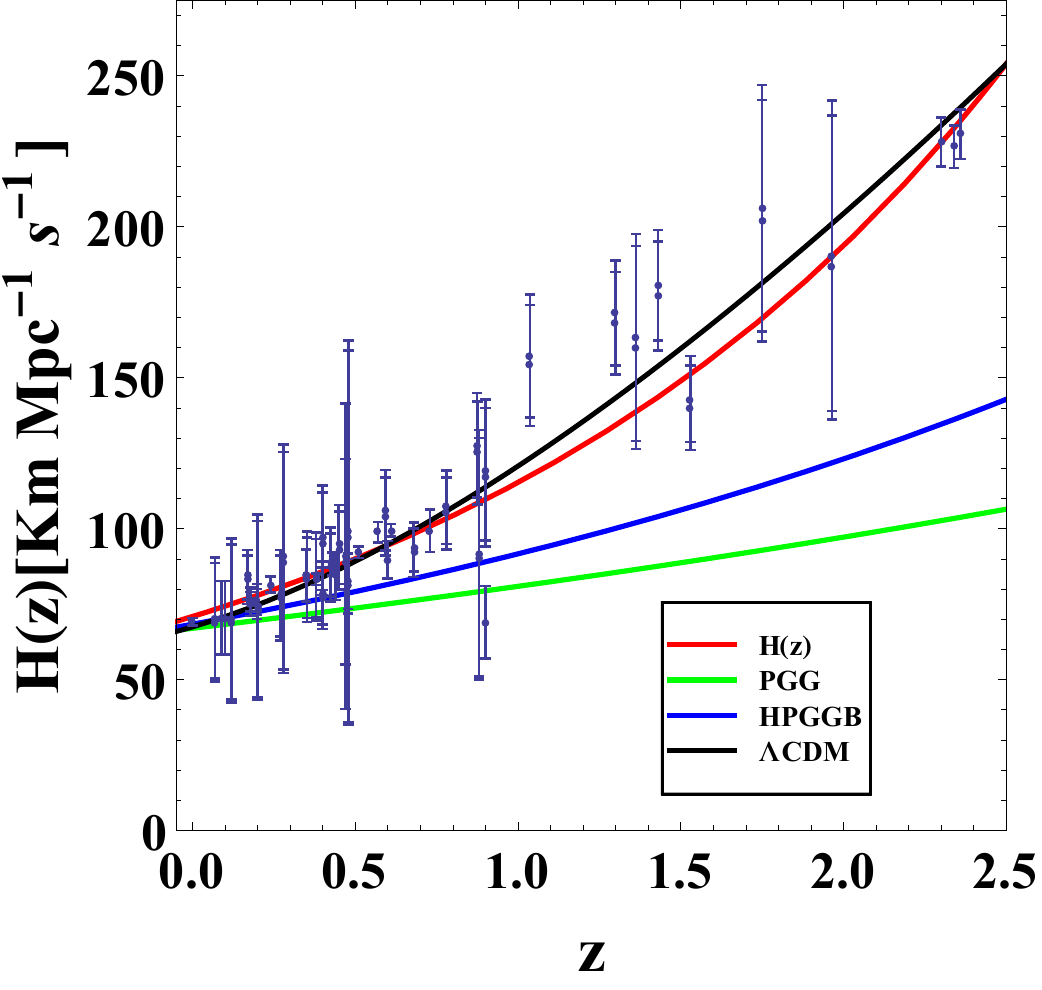}}\hfill
	\subfloat[]{\label{fig:2b}\includegraphics[scale=0.45]{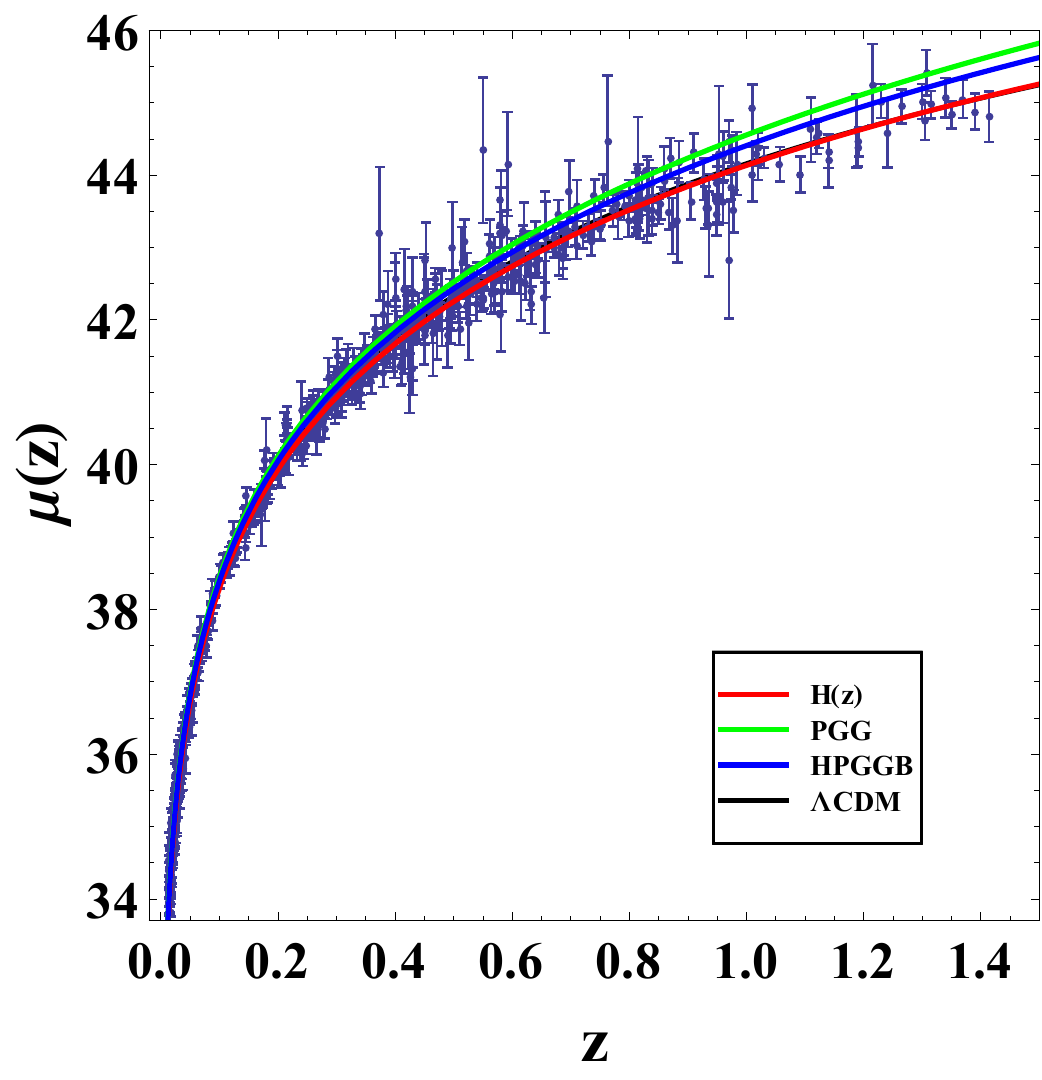}} 
\caption{The error bar plots for OHD and SNIa datasets show the likelihood of our model with $ \Lambda $CDM.}
 \label{fig:2}
\end{figure}

\subsection{BAO dataset}
We also use BAO, which are periodic fluctuations in the density of the visible baryonic matter of the Universe, to constrain our model parameters. Six points from the Baryon Acoustic Oscillation (BAO) dataset obtained from multiple surveys have been included in this analysis. These datasets include observations from the Sloan Digital Sky Survey (SDSS), the Baryon Oscillation Spectroscopic Survey (BOSS) LOWZ samples, and the Six Degree Field Galaxy Survey (6dFGS) \cite{Blake:2011en, SDSS:2009ocz}. According to \cite{SDSS:2005xqv}, the scale of dilation expression, $D_v(z)$, is as follows
\begin{equation}
    D_v(z)=\Big(\frac{d^2_A(z)z}{H(z)}\Big)^{1/3}.
\end{equation}
In this instance, the comoving angular diameter distance is denoted by $d_A(z)$, which has the following formal definition:
\begin{equation}
    d_A(z)=\int_0^z\frac{dz'}{H(z')}.
\end{equation}
The chi-squared function for the BAO analysis is defined as
\begin{equation}
    \chi^2_{BAO}=A^T C^{-1}_{BAO}A.
\end{equation}
$ A $ depends on the particular survey that is being examined, and $C^{-1}$ is the inverse of the covariance matrix \cite{{Giostri}}.

\subsection{Joint dataset}
\qquad For the joint dataset, we can minimize the value of $ \chi^2 $ by using the expression
\begin{equation}\label{21}
    \chi _{HPGGB}^{2}(\alpha,H_0)=\chi_{OHD}^{2}(\alpha,H_0)+\chi _{PGG}^{2}(\alpha,H_0)+\chi _{BAO}^{2}(\alpha,H_0).
\end{equation}

\begin{table}
\caption{ \bf Observational constraints on model parameters and $ H_0 $-tension consistency}
\begin{center}
\label{tabparm1}
\begin{tabular}{l c c c c c r} 
\hline\hline
\\ 
{ \bf {Dataset} } &  ~~~~~ $ \boldsymbol{ \alpha } $ &  ~~~~~  $ \boldsymbol{ H_0 } $ & {\bf $H_0$-tension consistency }   &   { \bf Reference}
\\
\\
\hline      
\\
{ OHD $ H(z) $  }     &  ~~~~~ $ 0.6636^{+0.0060}_{-0.0067} $   &  ~~~~~ $ 69.35 \pm 0.64 $  & P+BAO+BBN: $ 69.23 \pm 0.77$ & \cite{Chen:2021guo} ~
\\
\\
~&~ & ~& eBOSS+Planck mH2: $ 69.6 \pm 1.8 $ &  \cite{Pogosian:2020ded}
\\
\\
~&~ & ~& LSS teq standard ruler: $ 69.5^{+3.0}_{-3.5}$ &  \cite{Farren:2021jcd, Farren:2021grl}~
\\
\\
\hline      
\\
{PGG\footnote{Pantheon+Gold+Gamma-Ray Burst} } &  ~~~~~ $ 0.87 \pm 0.15 $   &  ~~~~~ $ 68.5 \pm 2.0 $   & P+Bispectrum+BAO+BBN: $ 68.3^{+0.83}_{-0.86} $  & \cite{Philcox:2021kcw, Philcox:2021tfv}~
\\
\\
~&~ & ~& WMAP9+BAO: $ 68.36^{+0.53}_{-0.52} $ &  \cite{Zhang:2019cww}~
\\
\\
~&~ & ~&  GW170817+VLBI: $ 68.3^{+4}_{-4.5} $ &  \cite{Mukherjee:2019qmm}~
\\
\\
\hline      
\\
{HPGGB\footnote{OHD+Pantheon+Gold+Gamma-Ray Burst+BAO} } &  ~~~~~$ 0.7992 \pm 0.0084 $  & ~~~~~ $ 68.0005 \pm 0.0094 $  & BOSS\footnote{BOSS correlation function}+BAO+BBN: $ 68.19\pm 0.99 $  & \cite{Zhang:2021yna}~ 
\\
\\  
~&~ & ~& BOSS DR12+BBN: $  67.9 \pm 1.1 $  &  \cite{Luo:2020dlg}~
\\
\\
~&~ & ~&  ACT CMB: $ 67.9 \pm 1.5 $  &  \cite{ACT:2020gnv}~ 
\\
\\
~&~ & ~&  CC, open $ w $CDM with systematics: $ 67.8^{+8.7}_{-7.2} $  & \cite{Moresco:2022phi}~ 
\\
\\
\hline\hline  
\end{tabular}    
\end{center}
\end{table}

This joint statistical study produces more substantial constraints on the model parameters. By checking the parameters constrained in Figs. \ref{fig:1b} and \ref{fig:1c}, we realize that the contours are closed but not in a proper oval shape. The convergence of these parameters can be checked through the Gelman-Rubin convergence test \cite{Singh:2022nfm}. This diagnostic test introduces $ \hat R $, the potential scale reduction, as a metric to gauge convergence for each model parameter. Here, $ \hat R=\sqrt{\frac{\rm Var(\theta)}{\rm W}} $, where $ \rm W $ is the variance within a chain and $ \rm Var(\theta) $ is the variance among the chains. Pioneering works \cite{Gelman:1992zz, Brooks:1998} suggest that $ \hat R $ values exceeding $ 1.2 $ for any model parameter are equivalent to a non-convergence. All the observed best-fit values of model parameters compatible with the Gelman-Rubin test are placed in Table \ref{tabparm1}. Now, to observe the similarity of our model with $ \Lambda $CDM, we draw error bar plots for OHD and SNIa datasets for the obtained best-fit values of $ \alpha $ and $ H_0 $ (see Fig. \ref{fig:2}).

\section{Physical consequences of the model}\label{sec:4}
\begin{figure}\centering
	\subfloat[]{\label{fig:3a}\includegraphics[scale=0.42]{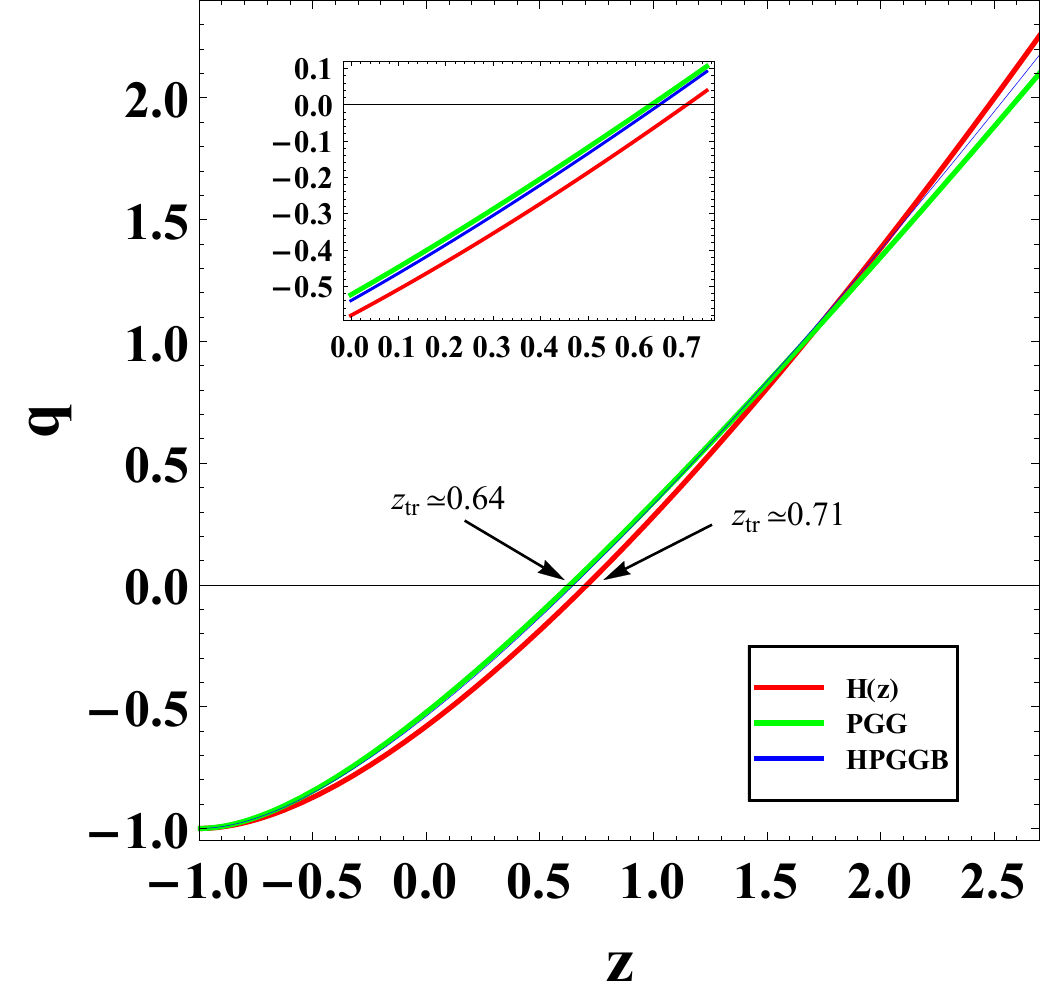}}\hfill
	\subfloat[]{\label{fig:3b}\includegraphics[scale=0.41]{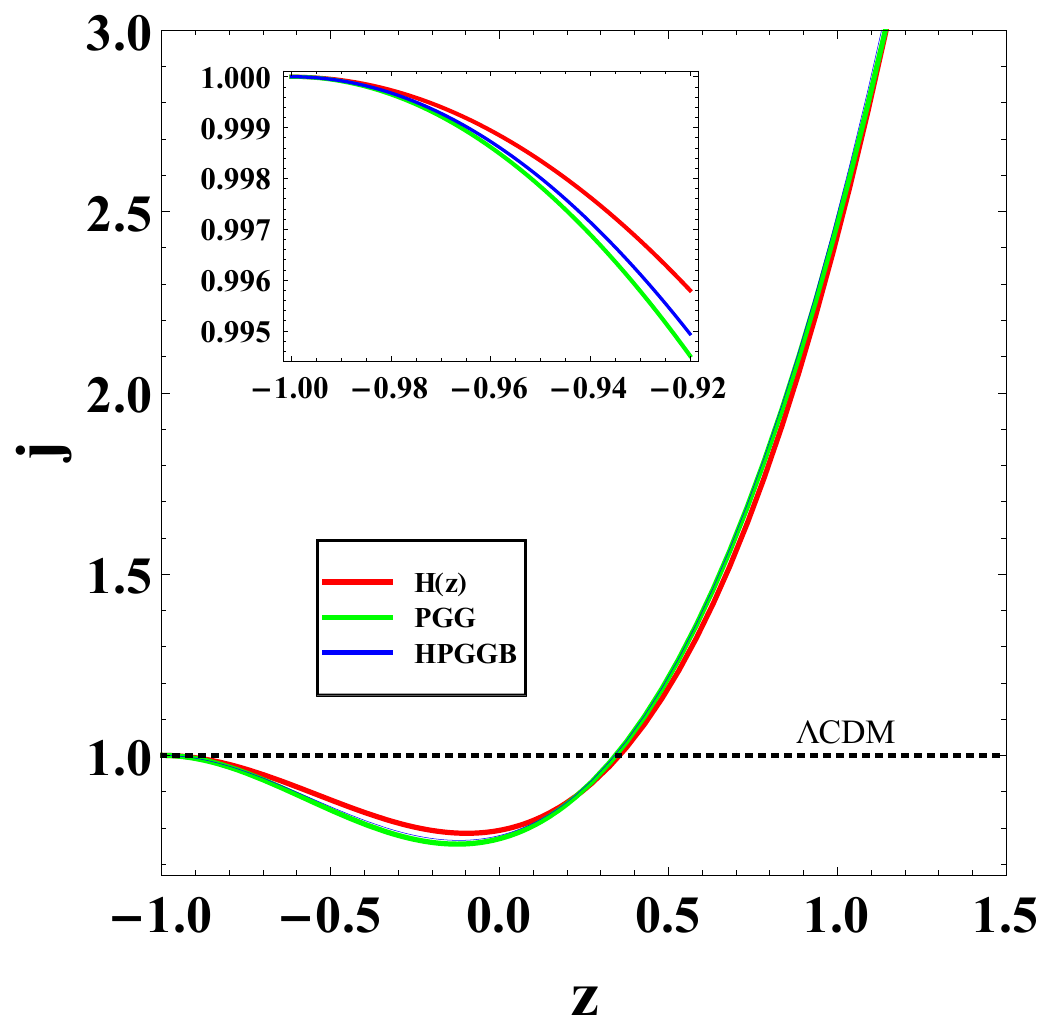}}\par
\caption{ The variations of the deceleration ($ q $) and the jerk ($ j $) parameters regarding the various observational datasets.}
 \label{fig:3}
\end{figure}

\qquad Over the last decades, cosmologists have grappled with deciphering the deceleration induced by gravitational forces. The deceleration parameter, used to measure the cosmic slowdown, was initially proposed without any inkling of the eventual revelation of the accelerated expansion of the Universe. According to recent affirmations, the expansion of the Universe is speeding up. To analyze the evolution of the deceleration parameter in a more significant way, we depicted the trajectories of $ q $ by using Eq. (\ref{13}) for all the mentioned datasets. In Fig. \ref{fig:3a}, $ z_{tr}$ represents the point of transition from deceleration to acceleration. In the late epochs, the Universe gracefully enters an accelerating de-Sitter regime, characterized by $ q=-1 $ \cite{Bolotin:2015dja}. Such behavior is compatible with the current dark energy-dominating era. 

Another set of parameters used to characterize cosmic evolution is built using higher-order derivatives of the scale factor, allowing constraints over different DE models \cite{, Alam:2003sc, Sahni:2002fz}. One of them is the jerk parameter $ j $, which is written as
\begin{equation}\label{22}
j=-q+2q(1+q)+(1+z)\frac{dq}{dz}.
\end{equation}

Using Eq. (\ref{13}), the explicit form of $ j $ is given by
\begin{equation}\label{23}
j=(\alpha +1)^{-2 \alpha } \left(2 \alpha ^{\alpha }-3 \alpha  (\alpha +1)^{\alpha }\right) \alpha ^{\alpha -2}+1,
\end{equation}
where $ j=1 $ recovers the $\Lambda $CDM model. In Fig. \ref{fig:3b}, we realize that $ j $ deviates from unity at the early and present epochs. Besides, it asymptotically converges to $ \Lambda $CDM at late times. Such a behavior is mapped using several distinct datasets.

\begin{figure}\centering
	\subfloat[]{\label{fig:4a}\includegraphics[scale=0.47]{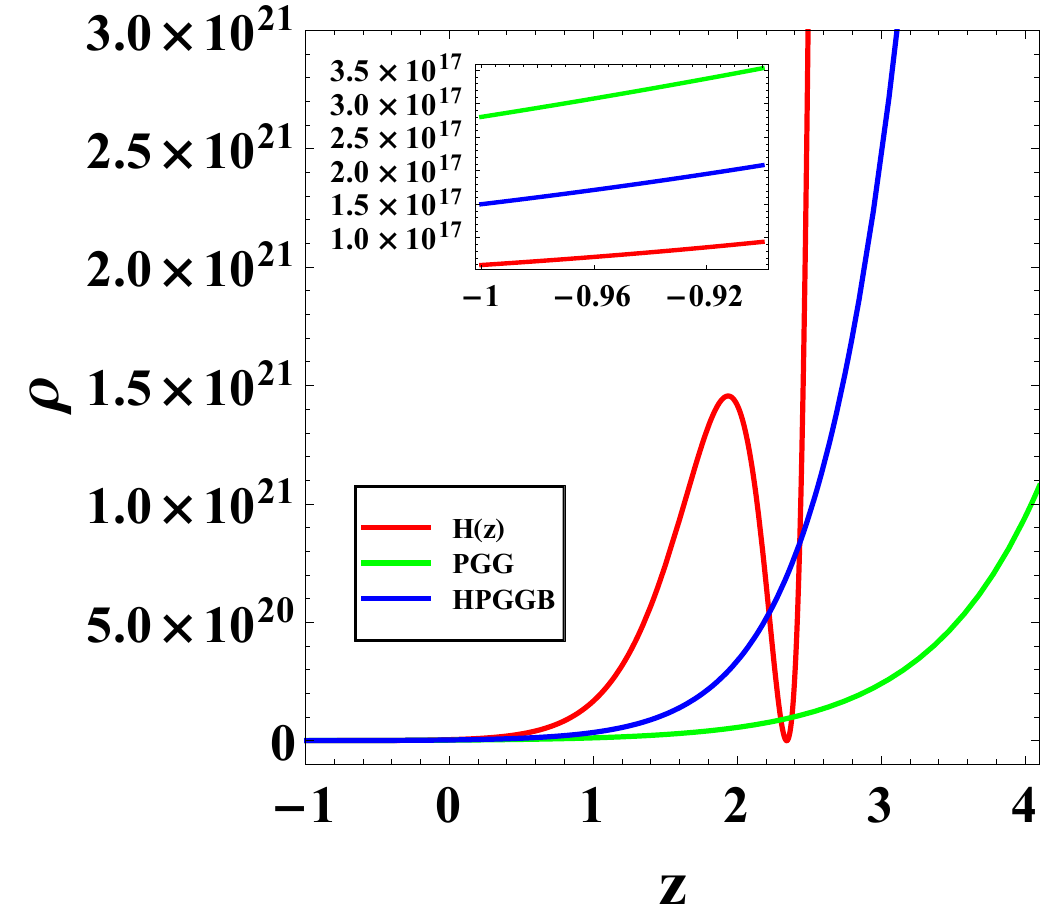}}	\hfill
	\subfloat[]{\label{fig:4b}\includegraphics[scale=0.47]{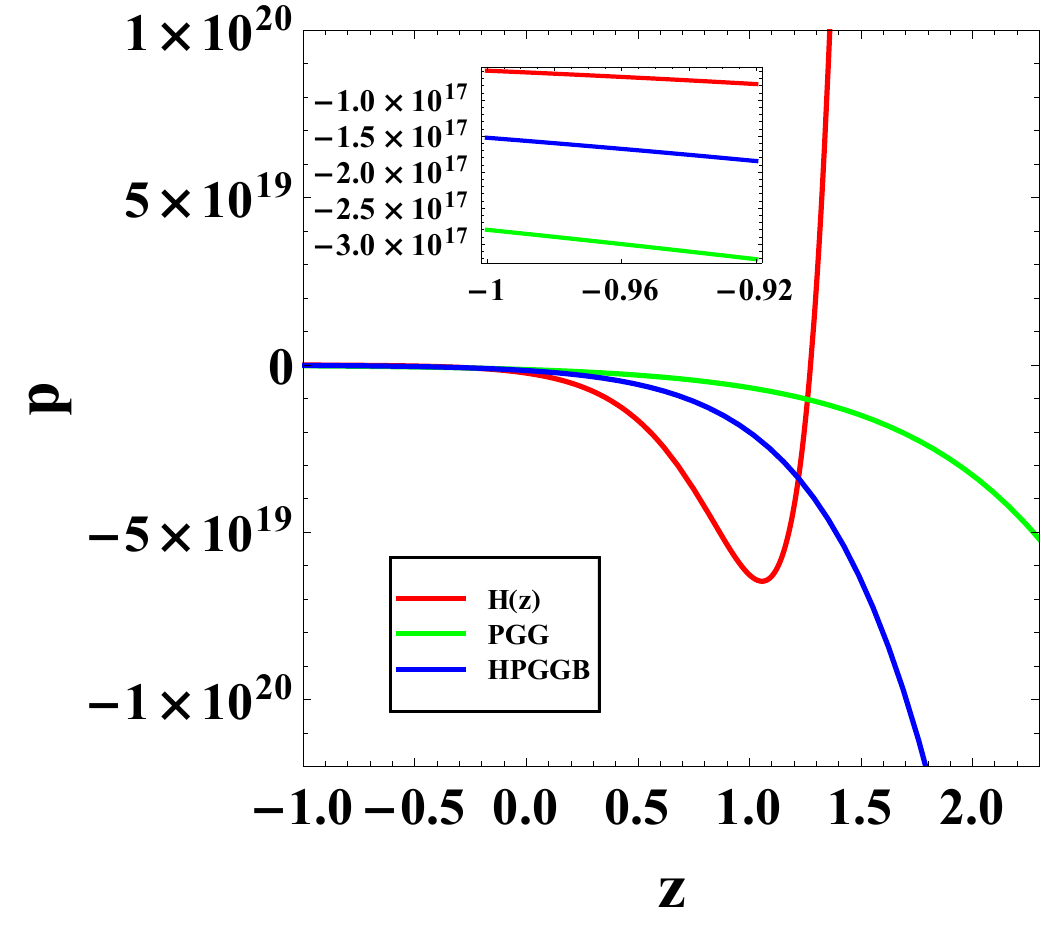}}\par
	\subfloat[]{\label{fig:4c}\includegraphics[scale=0.49]{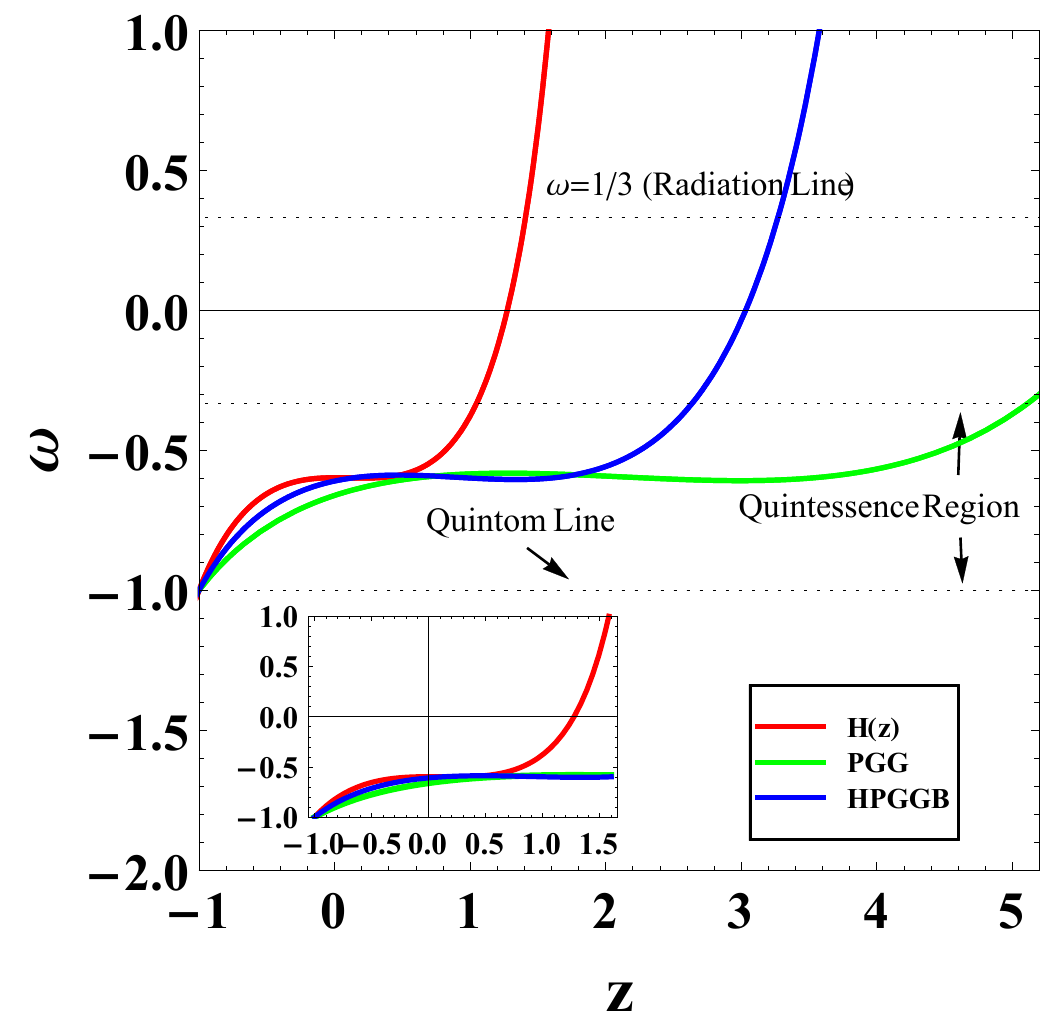}}
 \caption{ The variations of the energy density ($ \rho $), the isotropic pressure ($ p $), and the EoS ($ \omega $) parameters concerning the various observational datasets.}
\label{fig:4}	
\end{figure}

The analytic forms of $ \rho $ and $ p $ can be derived from Einstein field equations (\ref{11}), (\ref{12}) with $ \lambda=1 $ and $ n=0.5 $. As the expressions for $ \rho $ and $ p $ are lengthy, we plot these parameters directly instead of writing them. Fig. \ref{fig:4a} exhibits the evolution of energy density $ \rho $ from early to late times. Notably, $ \rho $ remains positive across the entire redshift range, decreasing to a positive value at present times. Moreover, Fig. \ref{fig:4b} provides a graphical description of pressure $ p $, whose values are negative at present, corroborating with an accelerating era. 

Now, let us carefully analyze the behavior of the equation of state (EoS) parameter $ \omega $. The EoS parameter is the ratio between pressure $ p $ and energy density $ \rho $. We illustrate its features for all observational datasets in Fig. \ref{fig:4c}. There, one can see that for higher redshift values, $ \omega $ is positive, which shows perfect fluid matter in the early times. Later, $ \omega $ enters the negative regime ($ -1<\omega<-1/3 $), compatible with a dark energy era. Moreover, we realize that $ \omega\to -1 $ as $ z\to -1 $, recovering the compatibility with the $ \Lambda $CDM model without accessing a phantom region.

\begin{figure}\centering
	\subfloat[]{\label{fig:5a}\includegraphics[scale=0.47]{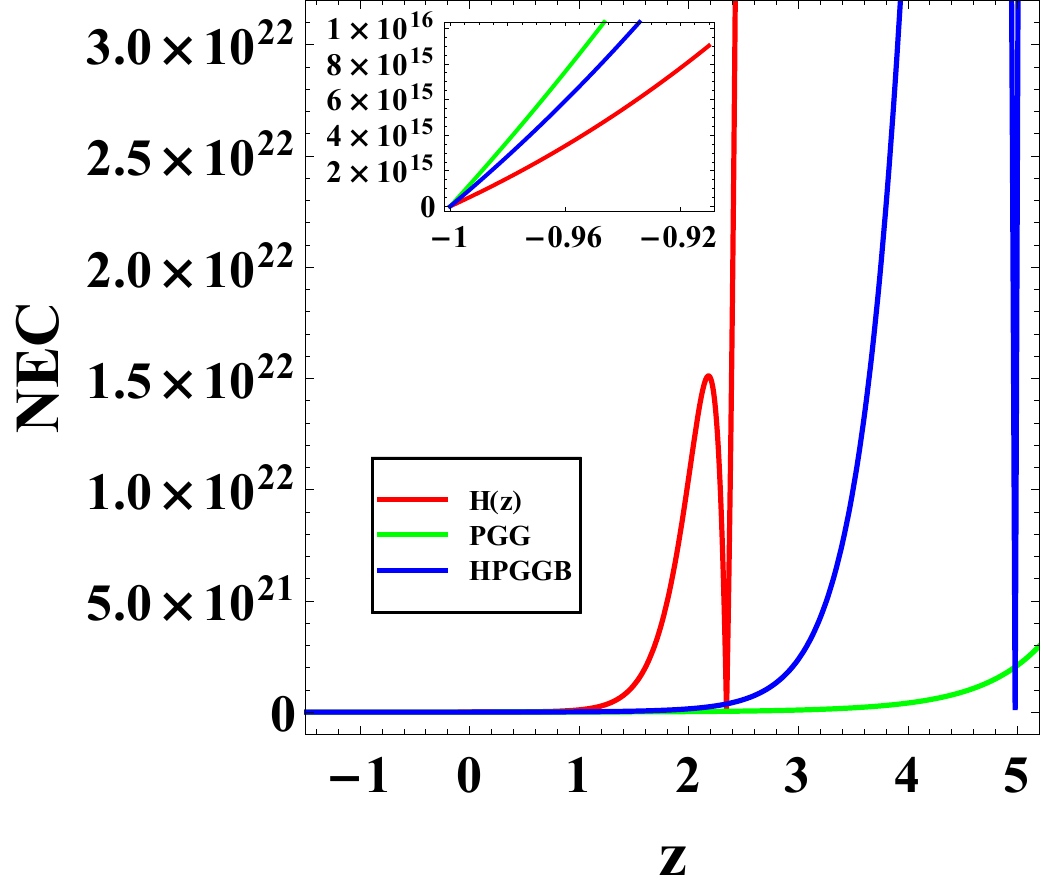}}	\hfill
	\subfloat[]{\label{fig:5b}\includegraphics[scale=0.45]{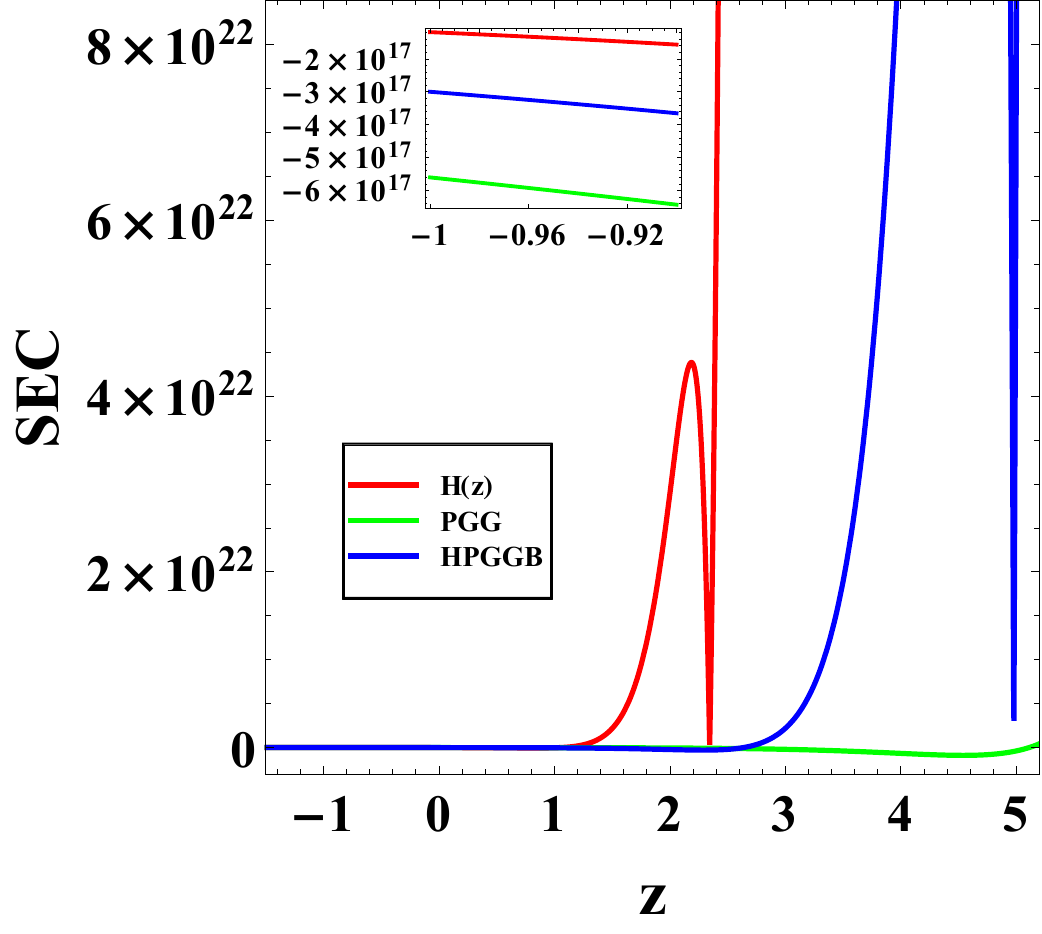}}\par
	\subfloat[]{\includegraphics[scale=0.49]{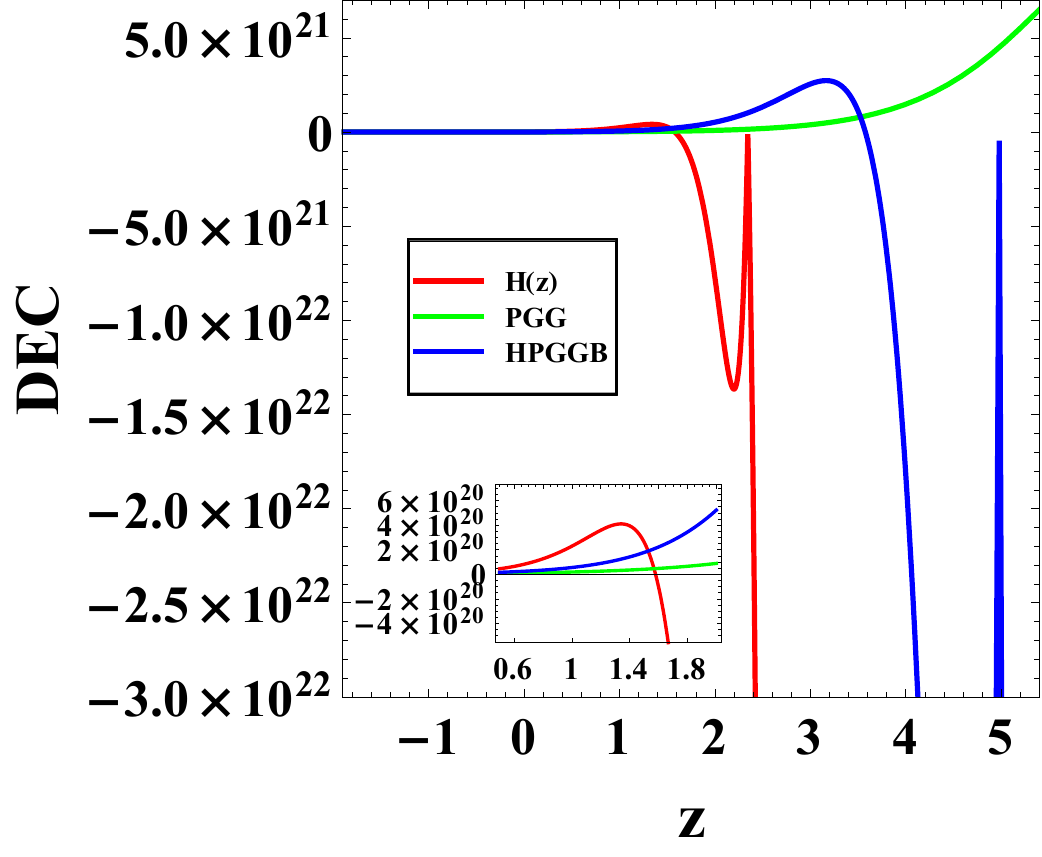}}
 \caption{ The NEC, SEC, and DEC plots concerning the various observational datasets.}
 \label{fig:5}
\end{figure}

An interesting investigation that we can perform as a matter of completeness is to verify the behavior of the energy conditions in this theory of gravity. The energy conditions are a set of constraints used to characterize the causal and geodesic structure of space-time. Such a set of conditions was introduced using the so-called Raychaudhuri equations \cite{Car}. The fundamental idea behind the energy conditions is that the whole energy-momentum tensor prohibits negative values for energy density \cite{Santos:2005pe, Santos:2007zza, Sen:2007ep}. In such an approach, we are going to work with the following set of energy conditions: $ \rho+p\geq 0 $ (Null Energy Condition $-$ NEC); $ \rho+3p\geq0 $ (Strong Energy Condition $-$ SEC); $ \rho>|p|\geq0 $ (Dominant Energy Condition $-$ DEC); $ \rho\geq0, ~~\rho+p\geq0 $ (Weak Energy Condition $-$ WEC).

Fig. \ref{fig:5} displays that NEC is satisfied for all the eras that our Universe passes through, corroborating with \cite{Qiu:2007fd}, but SEC and DEC are violated for specific ranges of $ z $. SEC is violated for small values of $ z $ for the datasets represented by red and blue curves in Fig. \ref{fig:5b}. While this EC is not satisfied for the range $ -1 \leq z < 6 $ for the dataset Pantheon+Gold+Gamma-Ray Burst (PGG). On the other hand, DEC is not satisfied for the redshift $ z > 1.55 $. Violation of strong energy condition implies the presence of exotic matter, which generates a repulsive force \cite{Singh:2019fpr}, and violation of dominant energy is connected with the presence of cosmological singularities and also with singularities related to black holes \cite{Lasukov:2020vxg}. Such behaviors for the energy conditions corroborate with scenarios where $\omega > 1/3$, surpassing the radiation era.

Another approach to check the similarities between various DE models and $ \Lambda $CDM model is a geometrical parameter pair technique titled statefinder diagnostic techniques (SDT) \cite{Singh:2015hva, Myrzakulov:2013owa, Rani:2014sia}. Since geometric parameters play a main role in exploring the evolution of the Universe, the statefinder diagnostic tool is one of the most important techniques to discuss. This tool was introduced by Alam et al. \cite{Alam:2003sc} and Sahni et al. \cite{ Sahni:2002yq, Sahni:2002fz}, where they defined the pair ($ s, r $), whose explicit forms are
\begin{equation}\label{24}
r=\frac{\dddot{a}}{aH^3}=-q+2q(1+q)+(1+z)\frac{dq}{dz},~~~~s=\frac{r-1}{3(q-\frac{1}{2})}, ~~~where ~~ q\neq \frac{1}{2}.  
\end{equation}

Then, by taking Eq. (\ref{13}), the previous equations are written as
\begin{equation}\label{25}
r=(\alpha +1)^{-2 \alpha } \left(2 \alpha ^{\alpha }-3 \alpha  (\alpha +1)^{\alpha }\right) \alpha ^{\alpha -2}+1, ~~~s=\frac{2}{3} \alpha ^{\alpha -1} (\alpha +1)^{-\alpha }.  
\end{equation}

\begin{figure}\centering
	\subfloat[]{\label{fig:6a}\includegraphics[scale=0.3]{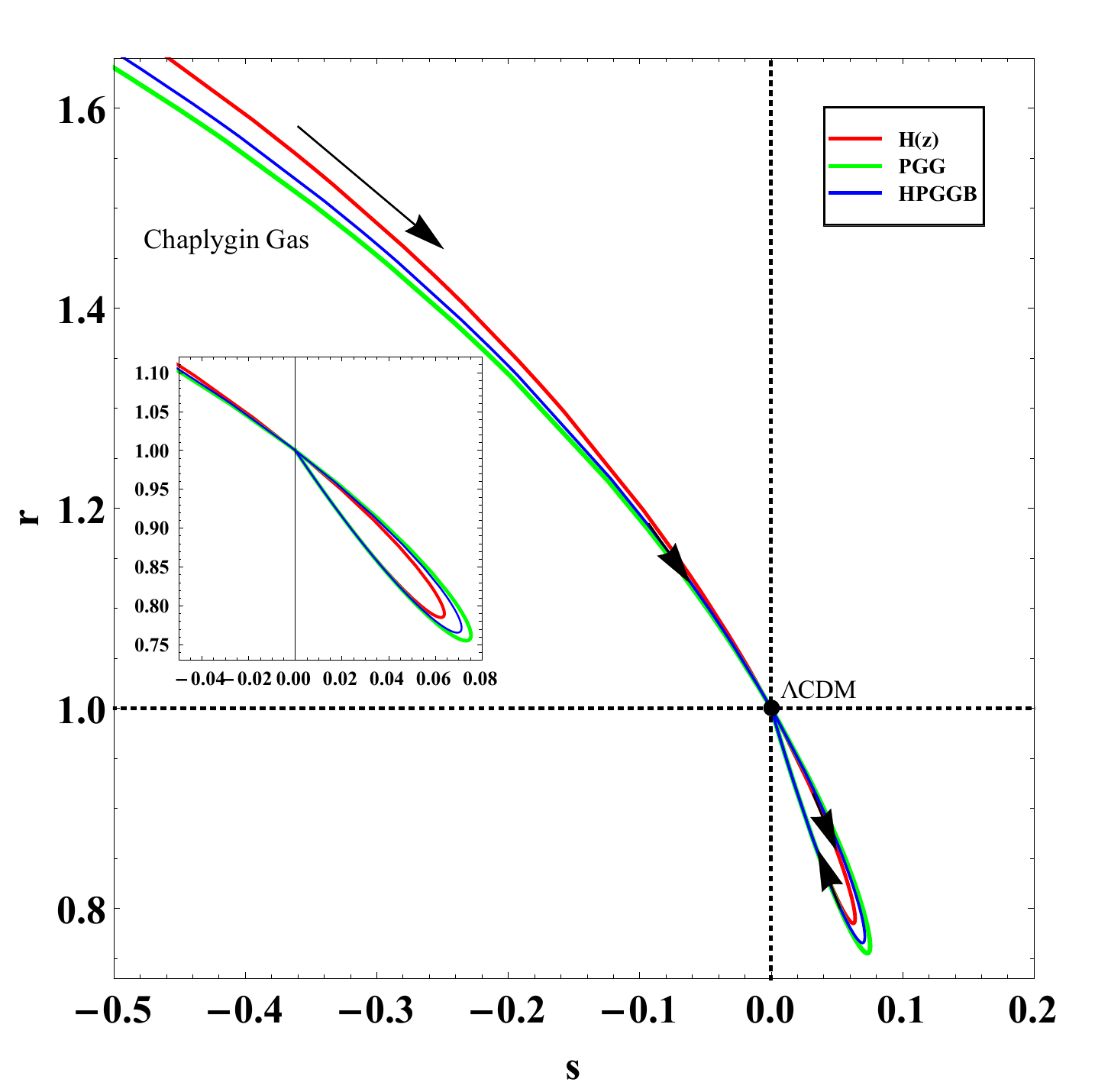}}\hfill
	\subfloat[]{\label{fig:6b}\includegraphics[scale=0.34]{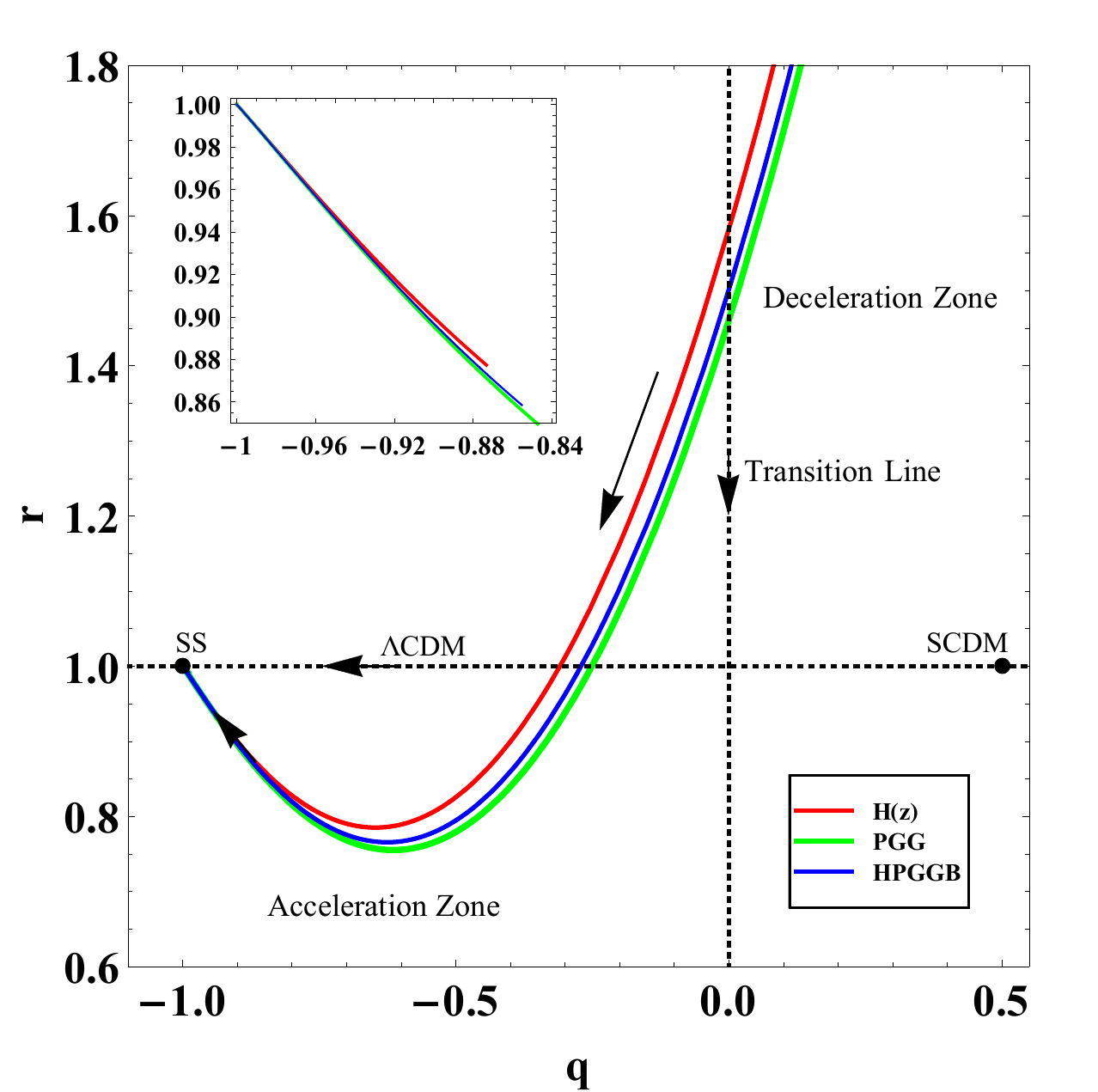}} 
\caption{ The statefinder plots for $ s-r $ and $ q-r $ regarding the various observational datasets.}
\label{fig:6}
\end{figure}

Here, we obtain both parameters as a function of $ \alpha $. Moreover, the parameter $ r $ is identical to the jerk parameter. One can use SD techniques to compare the DE models with the Holographic Dark Energy (HDE), Standard Cold Dark Matter (SCDM), and $ \Lambda $CDM models. The value of geometric parameters for the $ \Lambda $CDM model is $ s=0 $ and $ r=1 $; for the HDE model, $ s=2/3 $ and $ r=1 $; and for the SCDM model, $ s=1 $ and $ r=1 $. Thus, the $ s-r $ trajectories unveil the similarity of our model with the standard models. We depicted these trajectories for all the mentioned observational datasets using Eq. (\ref{25}). In Fig. \ref{fig:6a}, the arrows indicate the convergence of every $ s-r $ trajectory towards the $ \Lambda $CDM model. Each trajectory starts from the Chaplygin gas zone ($ r>1, s<0 $) and finally approaches the $ \Lambda $CDM model.

\begin{table}[htbp]
\caption{\bf The present values of the cosmological parameters}
\begin{center}\label{tabparm2}
\begin{tabular}{l c c c c r} 
\hline\hline
\\
{\bf{Dataset}} &  ~~~~~~~  $ \boldsymbol{ q } $ & ~~~~~ $ \boldsymbol{ j } $ & ~~~~~~  $ \boldsymbol{ \omega } $ & ~~~~~~ $ \boldsymbol{ z_{tr} } $\footnote{transition stage from deceleration to acceleration}
\\
\\
\hline 
\\
{OHD $ H(z) $}  &   ~~~~ $ -0.5811 $   &  ~~~~~ $ 0.7931  $ &  ~~~~~ $ -0.5974 $ &  ~~~~~  $ 0.7086 $ 
\\
\\
{Pan\footnote{Pantheon}+Gold+GRB\footnote{Gamma-Ray Burst} (PGG) }    & ~~~~ $ -0.5240 $   &  ~~~~~ $ 0.7701 $ &  ~~~~~ $ -0.6608  $ & ~~~~~ $ 0.6338 $ 
\\
\\
{OHD+Pan+Gold+GRB+BAO (HPGGB) }  &  ~~~~ $ -0.5407 $  & ~~~~~ $ 0.7791 $ &  ~~~~~ $ -0.6094  $ & ~~~~~ $ 0.6520 $   
\\ 
\\
\hline
\end{tabular}    
\end{center}
\end{table}

Moreover, the constraints over the pair ($ q,r $) are ($ \frac{1}{2},1 $) and ($ -1,1 $) for SCDM and Steady State (SS) models, respectively. Furthermore, the horizontal line $ r=1 $ corresponds to the $ \Lambda $CDM model. In Fig. \ref{fig:6b}, we can observe that the trajectories cross the $ \Lambda $CDM line for later times and converge to the SS model. This evolution of the trajectories occurs in the acceleration regime of our Universe, where $ q $ has negative values. 

\begin{table}
\caption{\bf Summary of the best-fit values for the Hubble parameter $ H_0 $ and the density parameter $ \Omega_{m0} $ of our model and the $\Lambda$CDM in $ f(R, L_m) $ gravity in comparison with the exponential $ F(R) $ gravity with logarithmic corrections, the standard exponential $ F(R) $ model, the $ F(R) $ model, and the $\Lambda$CDM model of Odintsov et al. \cite{Odintsov:2023cli, Odintsov:2024lid}.}
\begin{center}
\label{tabparm3}
\begin{tabular}{l l l l c c} 
\hline\hline
     \\ 
  {\bf Model}  &  ~~~~~{\bf Datasets}~~~~~~~~~~~~~~~~& ~~~~ $ \boldsymbol{H_0} $~~~~~~~~~~~~~~~~& ~~~~$ \boldsymbol{\Omega_{m0}} $
        \\
        \\
        \hline
        \\  
   $ f(R, L_m) $   &  ~~~~~$ H(z) $+Pan+Gold+GRB+BAO (HPGGB) &~~~ $ 68.0005 \pm 0.0094 $ &~~~$0.201^{+0.00012}_{-0.00017}$
      \\
       \\
           
  $\Lambda$CDM & ~~~~~$ H(z) $+Pan+Gold+GRB+BAO (HPGGB) ~~~~ &~~~ $ 68.00^{+0.014}_{-0.016} $  &~~~$0.198 \pm 0.0051$
       \\
       \\
         \hline
        \\  
   Exp $+\log F(R) $ & ~~~~~CC H(z)+SNeIa+CMB+BAO ~~~~ &~~~ $ 68.92^{+1.63}_{-1.72} $ &~~~$0.2984^{+0.0064}_{-0.0057}$  \cite{Odintsov:2024lid}
       \\
       \\
          Exp~$ +\log F(R)~ + $~axion  & ~~~~~CC H(z)+SNeIa+CMB+BAO ~~~~ &~~~ $ 69.0^{+1.72}_{-1.71} $ &~~~$0.2980^{+0.0065}_{-0.0065}$ \cite{ Odintsov:2024lid}
  \\
  \\
  Exp $ F(R)  $ & ~~~~~CC H(z)+SNeIa+CMB+BAO ~~~~ &~~~ $ 68.84^{+1.75}_{-1.64} $ &~~~$0.2913^{+0.0035}_{-0.0015}$ \cite{Odintsov:2024lid}
  \\
  \\
        \hline
        \\  
  $  F(R)~+ $ EDE   & ~~~~~CC H(z)+SNeIa+CMB+BAO ~~~~ &~~~ $ 68.93^{+1.61}_{-1.57} $ &~~~$0.294^{+0.0048}_{-0.0036}$ \cite{Odintsov:2023cli}
  \\
    \\
 $\Lambda$CDM ~~~~     & ~~~~~CC H(z)+SNeIa+CMB+BAO  & ~~~~$ 68.98^{+1.58}_{-1.60} $ &~~~$0.2908^{+0.0013}_{-0.0012}$ \cite{Odintsov:2023cli, Odintsov:2024lid} 
       \\
       \\
\hline\hline  
\end{tabular}   
\end{center}
\end{table}

\section{Conclusions} \label{sec:5}
This work carefully explored the cosmological constraints for the $ f(R, L_m) $ gravity, based on a flat Friedmann-Lemaître-Robertson-Walker (FLRW) space-time. By taking a parametrization involving the dimensionless parameter $ q $ and its correlation with the scale factor $ a $, we unveiled the transition from an early deceleration phase to the subsequent accelerated expansion of the Universe. We applied the MCMC method in the Hubble dataset $ H(z) $, Pantheon+Gold+Gamma-Ray Burst (PGG) dataset, and $ H(z) $+Pantheon+Gold+Gamma-Ray Burst+BAO (HPGGB) dataset (see Table \ref{tabparm1}), to constraint and evaluate the best-fit values of our free parameters ($ H_0 $ and $ \alpha $). The error bars depicted in the graphics of Fig. \ref{fig:2} show our model's higher degree of likeness with the $ \Lambda $CDM model at low redshift. The derived values for the present deceleration parameter, jerk parameter, and EoS parameter are highlighted in Table \ref{tabparm2} and corroborate with current observations. As redshift proceeds towards $-1$, the jerk parameter tends towards unity, recovering the $\Lambda$CDM-like behavior later, as shown in Fig. \ref{fig:3}.

The energy density $ \rho $ is positive for the whole redshift range and decreases with damped oscillations from high to low redshift, which, as far as we know, makes this work distinct from other papers in $ f(R, L_m) $ gravity. Initially, $ \rho $ is infinitely high, but it decreases at late times. The pressure $ p $ is positive during the Universe's early evolution, which indicates the structure formation phase, and later on, the pressure becomes negative, which shows the presence of a repulsive force resulting in the Universe's accelerating expansion. The EoS parameter $ \omega $ is supposed to transit from the initial singularity to nucleosynthesis, followed by structure formation until it reaches the dark energy-dominated era. As shown in Fig. \ref{fig:4c}, our model converges to $ \Lambda $CDM as $ z\to -1 $ for each observational dataset. Moreover, the graphics presented in Fig. \ref{fig:4} unveil that our model corroborates with an accelerated expanding quintessence model in later times.

In respect to the energy conditions, we found that the null energy condition is required to satisfy the EoS $ \omega\geq-1 $, which shows the matter and dark energy dominating nature of the Universe \cite{Qiu:2007fd} and this condition is satisfied in our approach. The strong energy condition is violated in late times, indicating the potential existence of exotic matter, which can be interpreted as the source for the accelerated phase of the Universe. The dominant energy condition is violated in the early evolution of the Universe, whereas it is satisfied in late times. Moreover, the $ s-r $ trajectories unveil that our model starts from the Chaplygin gas zone, then enters the quintessence region ($ r<1, s>0 $), and finally converges to the $ \Lambda $CDM model. The $ q-r $ trajectories transit from the decelerating phase to the accelerating phase, and after crossing the $ \Lambda $CDM threshold, they converge to the Steady State model (see Fig. \ref{fig:6}). In Table \ref{tabparm3}, we compare the best-fit values for the Hubble parameter $ H_0 $ and the density parameter $ \Omega_{m0} $ of our model and $\Lambda$CDM in $ f(R, L_m) $ gravity with the exponential $ F(R) $ gravity with logarithmic corrections, the standard exponential $ F(R) $ model, the $ F(R) $ model, and the $\Lambda$CDM model \cite{Odintsov:2023cli, Odintsov:2024lid}.

Therefore, our proposal for $ f(R, L_m) $ gravity, based on a coupling between matter and higher-order curvature with a distinctive form for the deceleration parameter, is capable of describing the evolution of different eras that our Universe passes through in a continuous way. The model is well constrained by observational data and nicely aligns with the $ \Lambda $CDM model. Also, we believe that the approach applied in our model may be helpful for further study in other theories of gravity like $ f(R, T)- \Lambda(\phi) $ \cite{Santos:2023zob}, $ f(R, T) $ \cite{Moraes:2016gpe}, $ f(Q) $ \cite{BeltranJimenez:2017tkd, Mandal:2020lyq}, $ f(Q, T) $ \cite{Xu:2019sbp, Arora:2020met}, and Braneworld cosmology \cite{Brax:2004xh}, etc.

\vskip0.2in

\section*{Acknowledgement}
JRLS would like to thank CNPq, Brazil (Grant 309494/2021-4), and PRONEX/CNPq/FAPESQ-PB (Grant nos. 165/2018 and 0015/2019) for financial support. The authors wish to extend their appreciation to the reviewer for his insightful criticisms, which have contributed to the current improvement of the paper.

\section*{Data Availability Statement} In this manuscript, we have used observational data as available in the literature as such our work does not produce any form of new data.

\section*{Conflicts of Interest} The authors assert that there are no conflicts of interest in this work's publication.

\section*{Declaration of Generative AI and AI-assisted technologies in the writing process} During the preparation of this work, the authors used ChatGPT by OpenAI to gain insights into writing minor pieces of the manuscript. The entire text was reviewed and rewritten properly by the authors, who take full responsibility for the publication's content.


\begin{thebibliography}{99}

\bibitem{SupernovaSearchTeam:1998fmf}
A.~G.~Riess \textit{et al.} [Supernova Search Team],
Astron. J. \textbf{116}, 1009-1038 (1998).

\bibitem{SupernovaCosmologyProject:1998vns}
S.~Perlmutter \textit{et al.} [Supernova Cosmology Project],
Astrophys. J. \textbf{517}, 565-586 (1999).

\bibitem{WMAP:2003xez}
E.~Komatsu \textit{et al.} [WMAP],
Astrophys. J. Suppl. \textbf{148}, 119-134 (2003).

\bibitem{Caldwell:2003hz}
R.~R.~Caldwell and M.~Doran,
Phys. Rev. D \textbf{69}, 103517 (2004).

\bibitem{2dFGRS:2005yhx}
S.~Cole \textit{et al.} [2dFGRS],
Mon. Not. Roy. Astron. Soc. \textbf{362}, 505-534 (2005).

\bibitem{Koivisto:2005mm}
T.~Koivisto and D.~F.~Mota,
Phys. Rev. D \textbf{73}, 083502 (2006).

\bibitem{Daniel:2008et}
S.~F.~Daniel, R.~R.~Caldwell, A.~Cooray and A.~Melchiorri,
Phys. Rev. D \textbf{77}, 103513 (2008).

\bibitem{SupernovaSearchTeam:2004lze}
A.~G.~Riess \textit{et al.} [Supernova Search Team],
Astrophys. J. \textbf{607}, 665-687 (2004).


\bibitem{WMAP:2003elm}
D.~N.~Spergel \textit{et al.} [WMAP],
Astrophys. J. Suppl. \textbf{148}, 175-194 (2003).

\bibitem{SDSS:2005xqv}
D.~J.~Eisenstein \textit{et al.} [SDSS],
Astrophys. J. \textbf{633}, 560-574 (2005).

\bibitem{SDSS:2009ocz}
W.~J.~Percival \textit{et al.} [SDSS],
Mon. Not. Roy. Astron. Soc. \textbf{401}, 2148-2168 (2010).

\bibitem{Jassal:2005qc}
H.~K.~Jassal, J.~S.~Bagla and T.~Padmanabhan,
Phys. Rev. D \textbf{72}, 103503 (2005).

\bibitem{Padmanabhan:2006ag}
T.~Padmanabhan,
AIP Conf. Proc. \textbf{861}, no.1, 179-196 (2006).

\bibitem{Copeland:2006wr}
E.~J.~Copeland, M.~Sami and S.~Tsujikawa,
Int. J. Mod. Phys. D \textbf{15}, 1753-1936 (2006).

\bibitem{Padmanabhan:2007xy}
T.~Padmanabhan,
Gen. Rel. Grav. \textbf{40}, 529-564 (2008).

\bibitem{Durrer:2007re}
R.~Durrer and R.~Maartens,
Gen. Rel. Grav. \textbf{40}, 301-328 (2008).

\bibitem{Capozziello:2011et}
S.~Capozziello and M.~De Laurentis,
Phys. Rept. \textbf{509}, 167-321 (2011).

\bibitem{Clifton:2011jh}
T.~Clifton, P.~G.~Ferreira, A.~Padilla and C.~Skordis,
Phys. Rept. \textbf{513}, 1-189 (2012).

\bibitem{Bamba:2012cp}
K.~Bamba, S.~Capozziello, S.~Nojiri and S.~D.~Odintsov,
Astrophys. Space Sci. \textbf{342}, 155-228 (2012).

\bibitem{Will:2014kxa}
C.~M.~Will,
Living Rev. Rel. \textbf{17}, 4 (2014).

\bibitem{Joyce:2014kja}
A.~Joyce, B.~Jain, J.~Khoury and M.~Trodden,
Phys. Rept. \textbf{568}, 1-98 (2015).

\bibitem{Cai:2015emx}
Y.~F.~Cai, S.~Capozziello, M.~De Laurentis and E.~N.~Saridakis,
Rept. Prog. Phys. \textbf{79}, no.10, 106901 (2016).

\bibitem{Bahamonde:2021gfp}
S.~Bahamonde, K.~F.~Dialektopoulos, C.~Escamilla-Rivera, G.~Farrugia, V.~Gakis, M.~Hendry, M.~Hohmann, J.~Levi Said, J.~Mifsud and E.~Di Valentino,
Rept. Prog. Phys. \textbf{86}, no.2, 026901 (2023).

\bibitem{Langlois:2018dxi}
D.~Langlois,
Int. J. Mod. Phys. D \textbf{28}, no.05, 1942006 (2019).

\bibitem{Frusciante:2019xia}
N.~Frusciante and L.~Perenon,
Phys. Rept. \textbf{857}, 1-63 (2020).

\bibitem{Arai:2022ilw}
S.~Arai, K.~Aoki, Y.~Chinone, R.~Kimura, T.~Kobayashi, H.~Miyatake, D.~Yamauchi, S.~Yokoyama, K.~Akitsu and T.~Hiramatsu, \textit{et al.}
PTEP \textbf{2023}, no.7, 072E01 (2023).

\bibitem{Odintsov:2023weg}
S.~D.~Odintsov, V.~K.~Oikonomou, I.~Giannakoudi, F.~P.~Fronimos and E.~C.~Lymperiadou,
Symmetry \textbf{15}, no.9, 1701 (2023).

\bibitem{Carroll:2000fy}
S.~M.~Carroll,
Living Rev. Rel. \textbf{4}, 1 (2001).

\bibitem{Peebles:2002gy}
P.~J.~E.~Peebles and B.~Ratra,
Rev. Mod. Phys. \textbf{75}, 559-606 (2003).

\bibitem{Carroll:2003wy}
S.~M.~Carroll, V.~Duvvuri, M.~Trodden and M.~S.~Turner,
Phys. Rev. D \textbf{70}, 043528 (2004).

\bibitem{Wetterich:1987fm}
C.~Wetterich,
Nucl. Phys. B \textbf{302}, 668-696 (1988).

\bibitem{Ratra:1987rm}
B.~Ratra and P.~J.~E.~Peebles,
Phys. Rev. D \textbf{37}, 3406 (1988).

\bibitem{Caldwell:1997ii}
R.~R.~Caldwell, R.~Dave and P.~J.~Steinhardt,
Phys. Rev. Lett. \textbf{80}, 1582-1585 (1998).

\bibitem{Armendariz-Picon:1999hyi}
C.~Armendariz-Picon, T.~Damour and V.~F.~Mukhanov,
Phys. Lett. B \textbf{458}, 209-218 (1999).

\bibitem{Armendariz-Picon:2000nqq}
C.~Armendariz-Picon, V.~F.~Mukhanov and P.~J.~Steinhardt,
Phys. Rev. Lett. \textbf{85}, 4438-4441 (2000).

\bibitem{Armendariz-Picon:2000ulo}
C.~Armendariz-Picon, V.~F.~Mukhanov and P.~J.~Steinhardt,
Phys. Rev. D \textbf{63}, 103510 (2001).

\bibitem{Mersini-Houghton:2001cwp}
L.~Mersini-Houghton, M.~Bastero-Gil and P.~Kanti,
Phys. Rev. D \textbf{64}, 043508 (2001).

\bibitem{Caldwell:1999ew}
R.~R.~Caldwell,
Phys. Lett. B \textbf{545}, 23-29 (2002).

\bibitem{Carroll:2003st}
S.~M.~Carroll, M.~Hoffman and M.~Trodden,
Phys. Rev. D \textbf{68}, 023509 (2003).

\bibitem{Sahni:1999gb}
V.~Sahni and A.~A.~Starobinsky,
Int. J. Mod. Phys. D \textbf{9}, 373-444 (2000).

\bibitem{Parker:1999td}
L.~Parker and A.~Raval,
Phys. Rev. D \textbf{60}, 063512 (1999).

\bibitem{Deffayet:2001pu}
C.~Deffayet, G.~R.~Dvali and G.~Gabadadze,
Phys. Rev. D \textbf{65}, 044023 (2002).

\bibitem{Freese:2002sq}
K.~Freese and M.~Lewis,
Phys. Lett. B \textbf{540}, 1-8 (2002).

\bibitem{Ahmed:2002mj}
M.~Ahmed, S.~Dodelson, P.~B.~Greene and R.~Sorkin,
Phys. Rev. D \textbf{69}, 103523 (2004).

\bibitem{Arkani-Hamed:2002ukf}
N.~Arkani-Hamed, S.~Dimopoulos, G.~Dvali and G.~Gabadadze,
[arXiv:hep-th/0209227 [hep-th]].

\bibitem{Dvali:2003rk}
G.~Dvali and M.~S.~Turner,
[arXiv:astro-ph/0301510 [astro-ph]].



\bibitem{deHaro:2023lbq}
J.~de Haro, S.~Nojiri, S.~D.~Odintsov, V.~K.~Oikonomou and S.~Pan,
Phys. Rept. \textbf{1034}, 1-114 (2023).

\bibitem{Singh:2018xjv}
J.~K.~Singh, K.~Bamba, R.~Nagpal and S.~K.~J.~Pacif,
Phys. Rev. D \textbf{97}, no.12, 123536 (2018).

\bibitem{Singh:2022jue}
J.~K.~Singh, H.~Balhara, K.~Bamba and J.~Jena,
JHEP \textbf{03}, 191 (2023)
[erratum: JHEP \textbf{04}, 049 (2023)].


\bibitem{Aviles:2014rma}
A.~Aviles, A.~Bravetti, S.~Capozziello and O.~Luongo,
Phys. Rev. D \textbf{90}, no.4, 043531 (2014).

\bibitem{delaCruz-Dombriz:2016bqh}
\'A.~de la Cruz-Dombriz, P.~K.~S.~Dunsby, O.~Luongo and L.~Reverberi,
JCAP \textbf{12}, 042 (2016).

\bibitem{Shaily:2024rjq}
Shaily, J.~K.~Singh and A.~Singh,
Fortsch. Phys. \textbf{72}, no.6, 2300244 (2024).

\bibitem{Capozziello:2019cav}
S.~Capozziello, R.~D'Agostino and O.~Luongo,
Int. J. Mod. Phys. D \textbf{28}, no.10, 1930016 (2019).


\bibitem{Shaily:2024nmy}
Shaily, A.~Singh, J.~K.~Singh and S.~Ray,
[arXiv:2402.01780 [gr-qc]].

\bibitem{Balhara:2023mgj}
H.~Balhara, J.~K.~Singh and E.~N.~Saridakis,
[arXiv:2312.17277 [gr-qc]].

\bibitem{Singh:2024ckh}
J.~K.~Singh, Shaily, H.~Balhara, S.~G.~Ghosh and S.~D.~Maharaj,
Phys. Dark Univ. \textbf{45}, 101513 (2024).

\bibitem{Shaily:2024xho}
Shaily, A.~Singh, J.~K.~Singh, S.~Hussain and R.~Myrzakulov,
[arXiv:2402.08709 [gr-qc]].


\bibitem{Nojiri:2007as}
S.~Nojiri and S.~D.~Odintsov,
Phys. Lett. B \textbf{657}, 238-245 (2007).

\bibitem{Nojiri:2007cq}
S.~Nojiri and S.~D.~Odintsov,
Phys. Rev. D \textbf{77}, 026007 (2008).

\bibitem{Cognola:2007zu}
G.~Cognola, E.~Elizalde, S.~Nojiri, S.~D.~Odintsov, L.~Sebastiani and S.~Zerbini,
Phys. Rev. D \textbf{77}, 046009 (2008).

\bibitem{Capozziello:2006ph}
S.~Capozziello, V.~F.~Cardone and A.~Troisi,
Mon. Not. Roy. Astron. Soc. \textbf{375}, 1423-1440 (2007).

\bibitem{Borowiec:2006qr}
A.~Borowiec, W.~Godlowski and M.~Szydlowski,
Int. J. Geom. Meth. Mod. Phys. \textbf{4}, 183-196 (2007).

\bibitem{Martins:2007uf}
C.~F.~Martins and P.~Salucci,
Mon. Not. Roy. Astron. Soc. \textbf{381}, 1103-1108 (2007).

\bibitem{Boehmer:2007kx}
C.~G.~Boehmer, T.~Harko and F.~S.~N.~Lobo,
Astropart. Phys. \textbf{29}, 386-392 (2008).

\bibitem{Sotiriou:2008rp}
T.~P.~Sotiriou and V.~Faraoni,
Rev. Mod. Phys. \textbf{82}, 451-497 (2010).

\bibitem{DeFelice:2010aj}
A.~De Felice and S.~Tsujikawa,
Living Rev. Rel. \textbf{13}, 3 (2010).

\bibitem{Bertolami:2007gv}
O.~Bertolami, C.~G.~Boehmer, T.~Harko and F.~S.~N.~Lobo,
Phys. Rev. D \textbf{75}, 104016 (2007).

\bibitem{Nojiri:2010wj}
S.~Nojiri and S.~D.~Odintsov,
Phys. Rept. \textbf{505}, 59-144 (2011).

\bibitem{Nojiri:2017ncd}
S.~Nojiri, S.~D.~Odintsov and V.~K.~Oikonomou,
Phys. Rept. \textbf{692}, 1-104 (2017).

\bibitem{Jana:2023djt}
B.~Jana, A.~Chatterjee, K.~Ravi and A.~Bandyopadhyay,
Class. Quant. Grav. \textbf{40}, no.19, 195023 (2023).

\bibitem{Maurya:2023lxt}
D.~C.~Maurya,
Phys. Dark Univ. \textbf{42}, 101373 (2023).

\bibitem{Pradhan:2022msm}
A.~Pradhan, D.~C.~Maurya, G.~K.~Goswami and A.~Beesham,
Int. J. Geom. Meth. Mod. Phys. \textbf{20}, no.06, 2350105 (2023).

\bibitem{Allemandi:2005qs}
G.~Allemandi, A.~Borowiec, M.~Francaviglia and S.~D.~Odintsov,
Phys. Rev. D \textbf{72}, 063505 (2005).

\bibitem{Nojiri:2004bi}
S.~Nojiri and S.~D.~Odintsov,
Phys. Lett. B \textbf{599}, 137-142 (2004).

\bibitem{Bertolami:2008zh}
O.~Bertolami, J.~Paramos, T.~Harko and F.~S.~N.~Lobo,
[arXiv:0811.2876 [gr-qc]].

\bibitem{Harko:2008qz}
T.~Harko,
Phys. Lett. B \textbf{669}, 376-379 (2008).

\bibitem{Bertolami:2009cd}
O.~Bertolami and M.~C.~Sequeira,
Phys. Rev. D \textbf{79}, 104010 (2009).

\bibitem{Solanki:2023onp}
R.~Solanki, Z.~Hassan and P.~K.~Sahoo,
Chin. J. Phys. \textbf{85}, 74-88 (2023).

\bibitem{Jaybhaye:2023lgr}
L.~V.~Jaybhaye, S.~Bhattacharjee and P.~K.~Sahoo,
Phys. Dark Univ. \textbf{40}, 101223 (2023).

\bibitem{Nojiri:2006ri}
S.~Nojiri and S.~D.~Odintsov,
Int. J. Geom. Meth. Mod. Phys. \textbf{4}, 115-146 (2007).

\bibitem{Singh:2024aml}
J.~K.~Singh, Shaily, H.~Balhara, K.~Bamba and J.~Jena,
Astron. Comput. \textbf{46}, 100790 (2024).

\bibitem{Kavya:2022dam}
N.~S.~Kavya, V.~Venkatesha, S.~Mandal and P.~K.~Sahoo,
Phys. Dark Univ. \textbf{38}, 101126 (2022).

\bibitem{Jaybhaye:2022gxq}
L.~V.~Jaybhaye, R.~Solanki, S.~Mandal and P.~K.~Sahoo,
Phys. Lett. B \textbf{831}, 137148 (2022).

\bibitem{Lobato:2021ehf}
R.~V.~Lobato, G.~A.~Carvalho and C.~A.~Bertulani,
Eur. Phys. J. C \textbf{81}, no.11, 1013 (2021).

\bibitem{Goncalves:2021vim}
B.~S.~Gon\c{c}alves, P.~H.~R.~S.~Moraes and B.~Mishra,
Fortsch. Phys. \textbf{71}, no.8-9, 2200153 (2023).

\bibitem{Harko:2010mv}
T.~Harko and F.~S.~N.~Lobo,
Eur. Phys. J. C \textbf{70}, 373-379 (2010).

\bibitem{Wang:2012rw}
J.~Wang and K.~Liao,
Class. Quant. Grav. \textbf{29}, 215016 (2012).

\bibitem{Singh:2022ptu}
J.~K.~Singh, Shaily, R.~Myrzakulov and H.~Balhara,
New Astron. \textbf{104}, 102070 (2023).

\bibitem{Banerjee:2005ef}
N.~Banerjee and S.~Das,
Gen. Rel. Grav. \textbf{37}, 1695-1703 (2005).

\bibitem{Shaily:2022enj}
Shaily, J.~K.~Singh, J.~R.~L.~Santos and M.~Zeyauddin,
Int. J. Mod. Phys. D \textbf{33}, 2450024 (2024).

\bibitem{Singh:2023ryd}
J.~K.~Singh, P.~Singh, E.~N.~Saridakis, S.~Myrzakul and H.~Balhara,
Universe \textbf{10}, 246 (2024).

\bibitem{Singh:2023gxd}
J.~K.~Singh, Shaily, A.~Singh, A.~Beesham and H.~Shabani,
Annals Phys. \textbf{455}, 169382 (2023).

\bibitem{Singh:2024kez}
J.~K.~Singh, H.~Balhara, Shaily and P.~Singh,
Astron. Comput. \textbf{46}, 100795 (2024).

\bibitem{Sharov:2018yvz}
G.~S.~Sharov and V.~O.~Vasiliev,
Math. Model. Geom. \textbf{6}, 1-20 (2018).

\bibitem{Pan-STARRS1:2017jku}
D.~M.~Scolnic \textit{et al.} [Pan-STARRS1],
Astrophys. J. \textbf{859}, no.2, 101 (2018).

\bibitem{Riess:1998dv}
A.~G.~Riess, R.~P.~Kirshner, B.~P.~Schmidt, S.~Jha, P.~Challis, P.~M.~Garnavich, A.~A.~Esin, C.~Carpenter, R.~Grashius and R.~E.~Schild, \textit{et al.}
Astron. J. \textbf{117}, 707-724 (1999).

\bibitem{Jha:2005jg}
S.~Jha, R.~P.~Kirshner, P.~Challis, P.~M.~Garnavich, T.~Matheson, A.~M.~Soderberg, G.~J.~M.~Graves, M.~Hicken, J.~F.~Alves and H.~G.~Arce, \textit{et al.}
Astron. J. \textbf{131}, 527-554 (2006).

\bibitem{Hicken:2009df}
M.~Hicken, P.~Challis, S.~Jha, R.~P.~Kirsher, T.~Matheson, M.~Modjaz, A.~Rest and W.~M.~Wood-Vasey,
Astrophys. J. \textbf{700}, 331-357 (2009).

\bibitem{Contreras:2009nt}
C.~Contreras, M.~Hamuy, M.~M.~Phillips, G.~Folatelli, N.~B.~Suntzeff, S.~E.~Persson, M.~Stritzinger, L.~Boldt, S.~Gonzalez and W.~Krzeminski, \textit{et al.}
Astron. J. \textbf{139}, 519-539 (2010).

\bibitem{SDSS:2014irn}
M.~Sako \textit{et al.} [SDSS],
Publ. Astron. Soc. Pac. \textbf{130}, no.988, 064002 (2018).

\bibitem{Izzo:2015vya}
L.~Izzo, M.~Muccino, E.~Zaninoni, L.~Amati and M.~Della Valle,
Astron. Astrophys. \textbf{582}, A115 (2015).

\bibitem{Demianski:2016dsa}
M.~Demianski, E.~Piedipalumbo, D.~Sawant and L.~Amati,
Astron. Astrophys. \textbf{598}, A113 (2017).

\bibitem{Amati:2018tso}
L.~Amati, R.~D'Agostino, O.~Luongo, M.~Muccino and M.~Tantalo,
Mon. Not. Roy. Astron. Soc. \textbf{486}, no.1, L46-L51 (2019).

\bibitem{Alcaniz:2002fy}
J.~S.~Alcaniz and J.~M.~F.~Maia,
Phys. Rev. D \textbf{67}, 043502 (2003).

\bibitem{bogna/2024}
B.~Szyk, Luminosity Calculator. Available at: https://www.omnicalculator.com/physics/luminosity. Accessed: Jun 24, 2024.

\bibitem{Blake:2011en}
C.~Blake, E.~Kazin, F.~Beutler, T.~Davis, D.~Parkinson, S.~Brough, M.~Colless, C.~Contreras, W.~Couch and S.~Croom, \textit{et al.}
Mon. Not. Roy. Astron. Soc. \textbf{418}, 1707-1724 (2011).

\bibitem{Giostri}
R. Giostri et al., 
J. Cosmol. Astropart. Phys. 2012(03) 027 (2012).

\bibitem{Singh:2022nfm}
J.~K.~Singh, Shaily, S.~Ram, J.~R.~L.~Santos and J.~A.~S.~Fortunato,
Int. J. Mod. Phys. D \textbf{32}, no.07, 2350040 (2023).

\bibitem{Chen:2021guo}
C.~Chen, C.~S.~Fischer, C.~D.~Roberts and J.~Segovia,
Phys. Rev. D \textbf{105}, no.9, 094022 (2022).

\bibitem{Pogosian:2020ded}
L.~Pogosian, G.~B.~Zhao and K.~Jedamzik,
Astrophys. J. Lett. \textbf{904}, no.2, L17 (2020).

\bibitem{Farren:2021jcd}
G.~S.~Farren, D.~Grin, A.~H.~Jaffe, R.~Hlo\v{z}ek and D.~J.~E.~Marsh,
Phys. Rev. D \textbf{105}, no.6, 063513 (2022).

\bibitem{Farren:2021grl}
G.~S.~Farren, O.~H.~E.~Philcox and B.~D.~Sherwin,
Phys. Rev. D \textbf{105}, no.6, 063503 (2022).

\bibitem{Philcox:2021kcw}
O.~H.~E.~Philcox and M.~M.~Ivanov,
Phys. Rev. D \textbf{105}, no.4, 043517 (2022).

\bibitem{Philcox:2021tfv}
O.~H.~E.~Philcox and Z.~Slepian,
Phys. Rev. D \textbf{103}, no.12, 123509 (2021).

\bibitem{Zhang:2019cww}
X.~Zhang and Q.~G.~Huang,
Sci. China Phys. Mech. Astron. \textbf{63}, no.9, 290402 (2020).

\bibitem{Mukherjee:2019qmm}
S.~Mukherjee, G.~Lavaux, F.~R.~Bouchet, J.~Jasche, B.~D.~Wandelt, S.~M.~Nissanke, F.~Leclercq and K.~Hotokezaka,
Astron. Astrophys. \textbf{646}, A65 (2021).

\bibitem{Zhang:2021yna}
P.~Zhang, G.~D'Amico, L.~Senatore, C.~Zhao and Y.~Cai,
JCAP \textbf{02}, no.02, 036 (2022).

\bibitem{Luo:2020dlg}
Y.~Luo, C.~Chen, M.~Kusakabe and T.~Kajino,
JCAP \textbf{05}, 042 (2021).

\bibitem{ACT:2020gnv}
S.~Aiola \textit{et al.} [ACT],
JCAP \textbf{12}, 047 (2020).

\bibitem{Moresco:2022phi}
M.~Moresco, L.~Amati, L.~Amendola, S.~Birrer, J.~P.~Blakeslee, M.~Cantiello, A.~Cimatti, J.~Darling, M.~Della Valle and M.~Fishbach, \textit{et al.}
Living Rev. Rel. \textbf{25}, no.1, 6 (2022).

\bibitem{Gelman:1992zz}
A.~Gelman and D.~B.~Rubin,
Statist. Sci. \textbf{7} 457-472 (1992).

\bibitem{Brooks:1998}
S. Brooks and A. Gelman,
J. Comp. Graph. Stat. \textbf{7}, no.4, 434-455 (1998).

\bibitem{Bolotin:2015dja}
Y.~L.~Bolotin, V.~A.~Cherkaskiy, O.~A.~Lemets, D.~A.~Yerokhin and L.~G.~Zazunov,
[arXiv:1502.00811 [gr-qc]].

\bibitem{Alam:2003sc}
U.~Alam, V.~Sahni, T.~D.~Saini and A.~A.~Starobinsky,
Mon. Not. Roy. Astron. Soc. \textbf{344}, 1057 (2003).

\bibitem{Sahni:2002fz}
V.~Sahni, T.~D.~Saini, A.~A.~Starobinsky and U.~Alam,
JETP Lett. \textbf{77}, 201-206 (2003).

\bibitem{Car}
S.~M.~Carroll, An Introduction to General Relativity: Spacetime and Geometry, San Francisco, CA, USA: Addison Wesley, (2004).

\bibitem{Santos:2005pe}
J.~Santos and J.~S.~Alcaniz,
Phys. Lett. B \textbf{619}, 11-16 (2005).

\bibitem{Santos:2007zza}
J.~Santos, J.~S.~Alcaniz, M.~J.~Reboucas and N.~Pires,
Phys. Rev. D \textbf{76}, 043519 (2007).

\bibitem{Sen:2007ep}
A.~A.~Sen and R.~J.~Scherrer,
Phys. Lett. B \textbf{659}, 457-461 (2008).

\bibitem{Qiu:2007fd}
T.~Qiu, Y.~F.~Cai and X.~M.~Zhang,
Mod. Phys. Lett. A \textbf{23}, 2787-2798 (2008).

\bibitem{Singh:2019fpr}
J.~K.~Singh and R.~Nagpal,
Eur. Phys. J. C \textbf{80}, no.4, 295 (2020).

\bibitem{Lasukov:2020vxg}
V.~Lasukov,
Symmetry \textbf{12}, no.3, 400 (2020).

\bibitem{Singh:2015hva}
J.~K.~Singh and S.~Rani,
Appl. Math. Comput. \textbf{259}, 187-197 (2015).

\bibitem{Myrzakulov:2013owa}
R.~Myrzakulov and M.~Shahalam,
JCAP \textbf{10}, 047 (2013).

\bibitem{Rani:2014sia}
S.~Rani, A.~Altaibayeva, M.~Shahalam, J.~K.~Singh and R.~Myrzakulov,
JCAP \textbf{03}, 031 (2015).

\bibitem{Sahni:2002yq}
V.~Sahni,
[arXiv:astro-ph/0211084 [astro-ph]].


\bibitem{Odintsov:2024lid}
S.~D.~Odintsov, D.~S\'aez-Chill\'on G\'omez and G.~S.~Sharov,
Phys. Dark Univ. \textbf{46}, 101558 (2024).


\bibitem{Odintsov:2023cli}
S.~D.~Odintsov, V.~K.~Oikonomou and G.~S.~Sharov,
Phys. Lett. B \textbf{843}, 137988 (2023).


\bibitem{Santos:2023zob}
J.~R.~L.~Santos, S.~S.~da Costa and R.~S.~Santos,
Phys. Dark Univ. \textbf{42}, 101356 (2023).

\bibitem{Moraes:2016gpe}
P.~H.~R.~S.~Moraes and J.~R.~L.~Santos,
Eur. Phys. J. C \textbf{76}, 60 (2016).

\bibitem{BeltranJimenez:2017tkd}
J.~Beltr\'an Jim\'enez, L.~Heisenberg and T.~Koivisto,
Phys. Rev. D \textbf{98}, no.4, 044048 (2018).

\bibitem{Mandal:2020lyq}
S.~Mandal, P.~K.~Sahoo and J.~R.~L.~Santos,
Phys. Rev. D \textbf{102}, no.2, 024057 (2020).

\bibitem{Xu:2019sbp}
Y.~Xu, G.~Li, T.~Harko and S.~D.~Liang,
Eur. Phys. J. C \textbf{79}, no.8, 708 (2019).

\bibitem{Arora:2020met}
S.~Arora, J.~R.~L.~Santos and P.~K.~Sahoo,
Phys. Dark Univ. \textbf{31}, 100790 (2021).

\bibitem{Brax:2004xh}
P.~Brax, C.~van de Bruck and A.~C.~Davis,
Rept. Prog. Phys. \textbf{67}, 2183-2232 (2004).

\end{thebibliography}
\end{document}